\documentclass[linenumbers]{aastex}

\usepackage[T1]{fontenc}
\usepackage{rotating}
\shorttitle{Cold gas-satellites correlation in TNG50}
\shortauthors{Damle et al.}
\graphicspath{{./}{figures/}}

\begin{document}
\nolinenumbers


\title{Searching for correlations between satellite galaxy populations and the cold circumgalactic medium around TNG50 galaxies}

\author[0000-0003-1386-5436]{Mitali Damle*}
\affiliation{New York University Abu Dhabi, Department of Physics, PO Box 129188, Abu Dhabi, UAE\\}
\affiliation{Center for Astrophysics and Space Science (CASS), New York University Abu Dhabi, PO Box 129188, Abu Dhabi, UAE}

\author[0000-0002-8710-9206]{Stephanie Tonnesen}
\affiliation{Center for Computational Astrophysics, Flatiron Institute, 162 5th Avenue, New York, NY 10010 USA}


\author[0000-0002-9735-3851]{Martin Sparre}
\affiliation{Institut für Physik und Astronomie, Universität Potsdam, Haus 28, Karl-Liebknecht Straße 24-25, 14476 Potsdam, Germany}
\affiliation{Leibniz-Institut für Astrophysik Potsdam (AIP), An der Sternwarte 16, 14482 Potsdam, Germany}

\author[0000-0002-1188-1435]{Philipp Richter}
\affiliation{Institut für Physik und Astronomie, Universität Potsdam, Haus 28, Karl-Liebknecht Straße 24-25, 14476 Potsdam, Germany}

\email{*mitali.damle@nyu.edu}



\begin{abstract}
We investigate the impact of satellites, a potentially important contributor towards the cold gas assembly of a halo, on the cold gas budgets of 197 TNG50 simulated halos with masses of 10$^{10.85}$ $\le$ M$_{200c}$/M$_{\odot}$ $\le$ 10$^{12.24}$ at $z$ = 0. To highlight the effect of satellites, we split the sample into three mass bins. We find that the total number of satellites, total mass of satellites, number of massive satellites and stellar mass of the most massive satellite, all correlate with the cold gas mass in halos. The total number of satellites (stellar mass of the most massive satellite) correlates most with the halo cold gas mass for low (middle) mass halos. The number of massive or observable satellites correlates with cold gas mass in similar manner as the total number of satellites. Our findings can, therefore, be used to guide future observers to focus on the link between the number of observable satellites and the amount of cold gas in a halo. Despite this correlation, we find that much of the cold gas lies far from the satellites. This leads us to conclude that satellites are unlikely to be the main supplier for cold gas in halos, however we discuss how they may act in tandem with other sources such that the satellite population correlates with the total cold gas in their host halo.

\end{abstract}



\section{Introduction} 
\label{sec:intro}

 The circumgalactic medium (CGM), or the gas within a dark matter halo but beyond the central galaxy’s own interstellar medium (ISM), is now considered to be a critical component of a galaxy’s baryon cycle \citep{tumlinson2017}. Importantly, observations have shown that the CGM contains a significant cold component (T $<$ 10$^5$ K), that is most likely to feed future star formation in the central galaxy. However, observations have not reached a clear consensus about the amount of mass in the cold CGM. While the few estimates for less massive, sub-Milky Way (MW) mass galaxies suggest a minor percentage (upto 3\%) of gas in the cold phase \citep{Barger_2016, Zheng_2024}, this number may increase substantially for more massive galaxies. 21-cm line estimates from high-velocity clouds (HVCs) in the MW \citep{Richter_2017} predict less than 10$^{9}$ M$_{\odot}$ (1-10$\%$ of the total CGM mass) in the cold CGM, while the low-$z$ universe COS-Halos observations \citep{Prochaska_2017} put this number at $\sim$ 9.2 $\pm$ 4.3 $\times$ 10$^{10}$ M$_{\odot}$ (up to 50$\%$ of the total CGM mass). Echoing this observational uncertainty, estimates from simulations \citep{hafen2019MNRAS.488.1248H,ramesh2023MNRAS.518.5754R} predict a broad range of cold gas mass fractions spanning the above mentioned mass percentage ranges obtained from observations. 
 
 This cold component has been observed at a range of radii within a variety of haloes. Recent observations showed high column densities of cold gas tracers like H I, Mg II and Fe II out to the virial radius in haloes with masses from 10$^{11}$ - 10$^{13}$ M$_{\odot}$ \citep{zhu2013a, chen2018MNRAS.479.2547C, lan2018ApJ...866...36L, zahedy2019MNRAS.484.2257Z}. Additionally, \citet{thom2012ApJ...758L..41T}, in their COS-Halos sample, find that the presence of cold CGM far from galaxies is not dependent on whether they are actively forming stars. 

  

There are many possible mechanisms that can impact the cool gas content of the CGM.  
In CGM gas where the ratio of radiative cooling time to free-fall time is low (t$_{\rm cool}/$t$_{\rm ff} <$ 10), thermally unstable perturbations can drive cooling of the hot CGM resulting in condensation into cold clouds (\citet{maller10.1111/j.1365-2966.2004.08349.x, sharma2012MNRAS.420.3174S,  voit2015Natur.519..203V, mccourt10.1093/mnras/stx2687, fielding2017impact, voit2019ambient} but see also \citet{esmerian2021MNRAS.505.1841E}). Several other processes can directly feed cold gas to the CGM as well as induce cooling of the hot CGM via perturbations and shocks or through mixing-induced cooling. For example, galactic outflows and fountain flows could export some cold gas from the inner regions of a galaxy into its CGM \citep{ford2013, faucher2015MNRAS.449..987F, faucher2016MNRAS.461L..32F, thompson2016MNRAS.455.1830T, liang2016MNRAS.458.1164L}, as well as cool their surroundings \citep{fraternali2015MNRAS.447L..70F}. Accretion of metal-poor, cold gas from the IGM could deposit cold gas into the CGM of galaxies on its way towards their ISM \citep{keres2005MNRAS.363....2K,Keres_2009,nelson2013MNRAS.429.3353N}. \citet{mandelker2020MNRAS.494.2641M} show that the mixing of cold gas flows with ambient hot medium is dictated by the relation between cooling time (growth of cool phase) and shear time (growth of warm phase). Minor mergers could also result in some cold gas from the smaller progenitor being dumped into the CGM of the main galaxy \citep{sancisi2008A&ARv..15..189S}. 

Satellite galaxies can also bring cold gas into the CGM \citep{suresh10.1093/mnras/sty3402}.  Within a more massive host halo, satellites experience numerous processes such as ram-pressure stripping \citep{gunn1972ApJ...176....1G}, tidal stripping \citep{bullock2001ApJ...548...33B}, strangulation \citep{larson1980ApJ...237..692L,bosch10.1111/j.1365-2966.2008.13230.x,peng2015Natur.521..192P}, and harassment \citep{moore1996Natur.379..613M,davies10.1093/mnras/sty3393} in addition to gas being ejected by feedback \citep{ostriker1999ApJ...513..252O, bernal2013ApJ...775...72B}. Observations indicate that satellites around Milky-Way like galaxies do indeed feed their gas into the CGM, as these satellites tend to exhibit shallower gas reservoirs, lower star formation rates and enhanced mass loss from their ISM/CGMs as opposed to isolated ones \citep{spekkens2014ApJ...795L...5S,Stierwalt_2015, fillingham2016MNRAS.463.1916F, Putman_2021}. In addition, many halo HI features seem to be associated with satellite galaxies (see \citet{putman2012} and references therein). Tidal streams arising from a satellite galaxy's motion through its host halo is one type of an extreme observational evidence in this regard. Detection of tidal streams around Whale galaxy \citep{Richter_2017} or the existence of the Magellanic Stream within our LG \citep{mastropietro2005MNRAS.363..509M, connors2006MNRAS.371..108C, richter2013ApJ...772..111R} are some excellent examples. Moreover, \citet{Rubin_2012} used absorption spectroscopy to argue that cool, enriched gas falling onto galaxies may be from surrounding dwarf satellites. 

Simulations find that a combination of removal and ejection results in gas flowing from satellites into the surrounding host CGM, and possibly onto the central host itself \citep{alcazar2017MNRAS.470.4698A, Sanchez_2018, Grand_2019, roy2023MNRAS.tmp.3042R}. In their HESTIA local group simulations, \citet{damle10.1093/mnras/stac663} find that many satellite galaxies have extended cold gas morphologies, indicating that a gas transfer from the satellites to the host CGM may occur. Simulated satellites can also induce localized cooling of gas in the extended CGM of the host \citep{nelson2020MNRAS.498.2391N,dutta10.1093/mnras/stab3653,Tonnesen2021, roy2023MNRAS.tmp.3042R}. 

Recently, \citet{Fielding_2020} compared the CGM properties in simulations spanning a broad range in resolutions and feedback models as well as whether they were isolated idealized halos or simulated in a cosmological context. They found that changing the stellar feedback models only strongly impacted the inner CGM properties.  Importantly, there was more cold gas mass in the outer (R $>$ 0.5 R$_{200}$\footnote{R$_{200}$ is the radius corresponding to the distance where the spherically averaged density equals 200 times the critical density of the Universe.}) CGM of galaxies that were simulated in cosmological volumes, indicating that accretion from gas flows and satellites may be important for the cold gas content of the CGM at large radii. 

 Our motivation for this study is to highlight the connection between the satellite populations and the cold gas in the CGM of sub-MW and MW-mass galaxies in TNG50 simulations. In particular, we aim to find which of the global satellite properties (number of satellites per halo, total mass in satellites, stellar mass of the most massive satellite or the number of massive satellites) shows the strongest correlation with the mass in the cold phase CGM within the final, $z$ = 0, (present-time) snapshot of the TNG50 simulation. 

This paper is structured as follows: We briefly describe the numerical setup and galaxy formation model of TNG50 in §\ref{section: Simulations} and establish the nomenclature (used throughout the paper) in §\ref{section: Definitions}. Sections §\ref{section: SampSelection} and §\ref{section: SatSampSelection} describe the sample selection process, in §\ref{section: BinSlice} we lay down the approach adopted for binning and slicing the sample. Section §\ref{section:CGM} describes the definition of cold gas in our simulations and our controlled experiment designed to bring out the satellite-CGM connection. We present our results in §\ref{section: Results}. The implications of our results in the context of current theories about CGM and galaxy formation and evolution are stated in §\ref{section: Discussion}. Finally, we sum up our conclusions and provide a quick discussion about certain caveats and ideas to be implemented in future projects (§\ref{section: Summary}).
\\

\section{Methods}
\label{section: Methods}

\subsection{Simulations}
\label{section: Simulations}

We use publicly available data from the TNG50 \citep{nelson2019ComAC...6....2N,nelson2019MNRAS.490.3234N,pillepich2019MNRAS.490.3196P} simulation. The TNG50 is the third simulation volume (the other two being TNG100 and TNG300) in the IllustrisTNG (hereafter, TNG) series \citep{nelson10.1093/mnras/stx3040,pillepich10.1093/mnras/stx3112,marinacci10.1093/mnras/sty2206,naiman10.1093/mnras/sty618,springel2018MNRAS.475..676S}. It is the smallest volume (box size = 51.7 cMpc$^3$), best resolved (m$_{\rm b}$= 8 $\times$ 10$^4$ M$_{\odot}$; m$_{\rm dm}$= 4.5 $\times$ 10$^5$ M$_{\odot}$) realization in the TNG suite and is generated on the underlying principles of combining the statistical prowess offered by conventional large box simulations with the superior resolution offered by zoom-in simulations. This approach aims to find a balance between the poor spatial resolution typical of large-volume simulations and the low number statistics that are a by-product of zoom-in simulations. 

The ideal MHD-self gravity coupled equations that form the backbone of TNG are solved using the moving-mesh code AREPO \citep{springel2010MNRAS.401..791S}. The tree-particle-mesh (TreePM) algorithm handles the Poisson equations for self-gravity while a second-order, finite-volume Godunov scheme \citep{pakmor2016MNRAS.455.1134P} is applied on the Voronoi mesh that continually moves in response to the underlying fluid motions. Baryon tracking is implemented via Monte-Carlo tracer particle scheme \citep{genel2013MNRAS.435.1426G} while halo catalogs and subsequent subhalo associations are computed via friends-of-friends (FoF) and SUBFIND algorithms \citep{springel2001} respectively.

The coupled interactions between dark matter, gas, stars and black hole particles are evolved from a redshift \textit{z} = 127 to present day (i.e. \textit{z} = 0). TNG uses a refined version of the galaxy formation model used in the original Illustris simulation \citep{torrey2014MNRAS.438.1985T,genel10.1093/mnras/stu1654,sijacki10.1093/mnras/stv1340,vogels2014MNRAS.444.1518V,vogelsberger2014Natur.509..177V}. The baryonic physics components unchanged from the original Illustris are: primordial and metal-line cooling \citep{vogelsberger2013MNRAS.436.3031V} along with a spatially-uniform UV background \citep{faucher2009} and self-shielding corrections in dense regions \citep{rahmati2013MNRAS.430.2427R}; star formation in the dense ISM implemented via stochastic sampling of particles satisfying a certain density threshold criterion; an effective equation of state for the two-phase model that enables energy transfer via radiative cooling and supernovae heating, ultimately resulting in a pressurized ISM \citep{springel2003MNRAS.339..289S}; chemical enrichment and mass loss from mono-age stellar populations; supermassive blackholes' seeding and quasar-mode feedback contributions. The novel features in the TNG model include an updated kinetic feedback model from supermassive blackholes (alongside the existing quasar-mode feedback), a revised galactic winds' prescription and inclusion of a seed magnetic field and its amplification. Interested readers can find further details on the TNG model in the relevant method papers \citep{weinberger2017MNRAS.465.3291W,pillepich2018MNRAS.473.4077P}.

The initial conditions assume the best-fit cosmological parameters ($h = 0.6774$, $\sigma_{8} = 0.8159$, n$_{s}$ = 0.9667, $\Omega_{\Lambda,0} = 0.6911$, $\Omega_{m,0} = 0.3089$ and $\Omega_{b,0} = 0.0486$) derived from the Planck 2016 \citep{planck2016A&A...594A..13P} results. 

\subsection{Object definitions}
\label{section: Definitions}

Before stating details about our sample selection and curation, we first describe in detail the relevant nomenclature and the TNG particle association scheme used throughout this paper. 

\begin{itemize}
    \item All halo groups, identified in TNG using the FoF algorithm with a linking length of \textit{b} = 0.2, are referred to as \textit{Halos}. 
    \item All galaxies, identified in TNG using the SUBFIND algorithm, are referred to as Subhalos. Each halo consists of a most massive subhalo plus a number of secondary subhalos. 
    \item The most massive subhalo of each respective FoF group is referred to as the \textit{Central}. 
    \item Secondary subhalos within a FoF group are referred to as \textit{Satellites}. Each satellite population consists of a \textit{most massive satellite}.
    \item There exist primarily four types of particles/cells in TNG: Gas (PartType0), Dark matter (PartType1), Stars and wind particles (PartType4) and Black holes (PartType5). We refer to the masses associated with each of these particle types corresponding to a single halo as M$_{\rm gas}$, M$_{h}$, M$_{*}$, and M$_{\rm bh}$ respectively.
    \item Each particle has a FoF halo\footnote{Particles that are not associated with \textit{any} FoF halo are referred to as \textit{Outer fuzz} particles.} and SUBFIND association\footnote{SUBFIND identifies substructures as locally dense, gravitationally bound groups of particles. It estimates local density for each FoF gas particle using adaptive kernel estimation with a set number of smoothing neighbors. This allows SUBFIND to decide whether this particular FoF particle is associated/bound to a subhalo or not.}. There also exists a (small) set of particles that are, by virtue of their binding energy, not associated with any subhalo but are instead only associated with the FoF halo. These particles are called the \textit{Inner fuzz}.
\end{itemize}


\subsection{Central Sample Selection}
\label{section: SampSelection}
Our goal is to ultimately find out if there exists any correlation between the properties of the overall satellite population to the cold gas phase of its sub-MW and MW-mass parent central at $z$ = 0. 

The gravitationally-bound dark matter-only mass (M$_{h}$) is close but not exactly equal to the total halo mass. The virial-overdensity mass (i.e. the mass enclosed in a sphere whose mean density is 200 times the critical density of the Universe, at the time the halo is considered), M$_{\rm 200c}$, generally represents the total halo mass more accurately. Hence, we adopt M$_{\rm 200c}$\footnote{We calculate this quantity for each halo by first counting the total number of dark matter particles within R$_{200}$ and then multiplying it with the mass of a TNG dark matter particle.} instead of M$_{h}$. Thereby, we adopt a set of constraints for the values of M$_{\rm 200c}$, M$_{*}$ and the star formation rate (SFR) and select centrals that meet all the three criterion. The ranges for the three aforementioned parameters are listed below-- 

\begin{itemize}
    \item M$_{\rm 200c}$: $\{7-175\}$ $\times$ 10$^{10}$ M$_{\odot}$  (or $\{10-250\}$ $\times$ 10$^{10}$ M$_{\odot}/h$)
    \item M$_{*}$: $\{0.7-7\}$ $\times$ 10$^{10}$ M$_{\odot}$  (or $\{1-10)\}$ $\times$ 10$^{10}$ M$_{\odot}/h$)
    \item SFR: $\{0.7-3.5\}$ M$_{\odot}$ yr$^{-1}$ 
\end{itemize}
A total of 234 parent FoF groups or halos are found to contain centrals that qualify in this selection process. 

\subsection{Satellite Sample Selection}
\label{section: SatSampSelection}

Each of the 234 selected centrals have several satellite galaxies associated with them. For this work we define associated satellites to be those within the central galaxy's R$_{200}$ at $z$=0, and do not impose any criteria on whether the satellites are gravitationally bound to the host halo. 

In TNG, there exist several stellar/dark matter overdensities resembling satellite galaxies that have a non-cosmological origin. In order to exclude these, we only consider subhaloes satisfying the SubhaloFlag = 1 criterion\footnote{See Sec. 5.2 in \citet{nelson2019ComAC...6....2N}.}. Since we want to include as many satellites with realistic particle masses as possible, we impose a minimum stellar mass cut of M$_{*}$ $\gtrsim$ 10$^5$ M$_{\odot}$ (in accordance with TNG50's baryonic mass resolution) on all satellites. 

We discover that some satellites have an unusually high M$_{*}$/M$_{h}$ ratio (typically exceeding unity) as compared to both the stellar-to-halo-mass (SHMR) relation for centrals in \citet{behroozi2013ApJ...770...57B} and that seen for TNG galaxies in \citet{engler2021MNRAS.500.3957E}. We regard these as `outliers' and they amount to 5.174\% of the total identified satellites. We verify that most of the outliers lie rather close to the centrals, within  $r <$ 0.2 R$_{200}$. \citet{engler2021MNRAS.500.3957E} demonstrate that satellite galaxies' SHMR is generally bound to be different from analog centrals due to the effects of tidal stripping on their dark matter material. In other words, dark matter gets stripped off to a greater extent than the centrally-concentrated baryonic matter. This results in an increase in the M$_{*}$/M$_{h}$ ratio, as seen for the outliers in our case. The presence of outlier satellites in our sample is, therefore, not surprising and we choose to retain them in our subsequent analysis\footnote{We verify that the inclusion or exclusion of outlier satellites has negligible qualitative implication on our subsequent results, as one would expect given that they are just 5\% of our total satellite sample.}.

\subsection{Binning \& Slicing}
\label{section: BinSlice}

In §\ref{section: SampSelection}, we described our sample selection that resulted in 234 sub-MW and MW-mass halos that span our defined ranges in halo and stellar masses as well as SFR. Our focus in this paper is to understand how the CGM mass in the cold gas phase (M$_{\rm cg}$) is affected by different properties of the satellite population. However, to do this we must first address the positive correlation between the number of satellites per halo, N$_{\rm sats}$ and the stellar masses of the central host galaxy. Massive halos with larger stellar masses generally harbor larger reserves of cold gas as well as more satellite halos \citep{munshi2013ApJ...766...56M,Feldmann_2019,henriques10.1093/mnras/stz577}. The interdependence between CGM cold gas mass and central galaxy stellar masses can cloud the more subtle scaling of cold gas mass with satellite properties.

To reduce this effect, we bin the sample by the central galaxy's stellar mass such that within a given bin, the N$_{\rm sats}$ does not correspond to an increase in the stellar mass of the central. As discussed in detail in Appendix §\ref{section: binning}, we define our mass bins by selecting stellar mass ranges of Centrals (M$_{\rm *C}$) whose mass distribution is not dependent on the number of satellites. Briefly, we use the 2-sample Kolmogorov-Smirnov (K-S) test to compare the distributions of the central galaxy stellar masses of a satellite-poor subsample (the bottom quartile of N$_{\rm sats}$ in the mass range) to a satellite-rich subsample (the top quartile of N$_{\rm sats}$ in the mass range) of central galaxies. We choose the largest mass ranges that give p-values above 0.05, indicating that the stellar mass distribution does not differ based on whether the galaxies are satellite-poor or satellite-rich. 
Using this approach, our mass bins are log M$_{\rm *C}$ = 10.160-10.405, log M$_{\rm *C}$ = 10.453-10.715 and log M$_{\rm *C}$ = 10.931-11.288.
Table~\ref{table:FinalSamp} shows our final sample constituting a total of 197 halos: 115 halos in log M$_{\rm *C}$ = 10.160-10.405, 61 halos in log M$_{\rm *C}$ = 10.453-10.715 and 21 halos in log M$_{\rm *C}$ = 10.931-11.288 bin\footnote{The remaining 37 from our original sample of 234 halos fall on either side of the log M$_{\rm *C}$ = 10.453-10.715 mass bin. Since including them in any of our mass bins results in a low p-value from the 2-sample KS test, we exclude them from our final sample.}. Further details of the binning process are explained in Appendix §\ref{section: binning}.

We note that the halo mass ranges (obtained from M$_{\rm 200c}$ values from Figs. §\ref{fig:LMBScatter}, §\ref{fig:MMBScatter} and §\ref{fig:HMBScatter}) for the first two bins (log M$_{\rm 200c} \leq$ 12.0) roughly correspond to less massive galaxies (i.e. sub-MW mass) while the ranges in the last bin (log M$_{\rm 200c} \geq$ 12.0) broadly align with more massive galaxies (i.e. MW-mass; \citet{posti2019A&A...621A..56P}). To further orient the reader, the halo mass values of some of the lowest mass sub-MW galaxies lie in the mass ranges of the Large Magellanic Cloud (LMC: log M$_{\rm 200c} \sim$ 11.3; \citet{fushimi2024A&A...688A.147F}). 



\begin{table*}
\centering
\begin{tabular}{cccc}
\hline
\hline

 & & Number of halos in log M$_{\rm *C}$ \\ Quartiles & 10.160-10.405 & 10.453-10.715 & 10.931-11.288\\
\hline
\hline
Bottom quartile (Satellite-poor) & 29 & 15 & 5\\
Inter-quartile & 57 & 31 & 11\\
Top quartile (Satellite-rich) & 29 & 15 & 5\\
\hline
\hline\\
\end{tabular}
\caption{Number of halos (in log M$_{\rm *C}$) after binning our sample. The three bins (10.160-10.415, 10.453-10.715 and 10.931-11.288) are computed on basis of the  M$_{\rm *C}$ values while the quartiles are calculated based on the total number of satellites per halo. The sum of halos across all the bin and quartiles is equal to our total sample size of 234 halos.}
\label{table:FinalSamp}
\end{table*}

\subsection{Defining the CGM around the central galaxy sample}
\label{section:CGM}

We use the gas particle spatial information as our main criteria for defining the CGM around our central galaxy sample. We fix the particle distances, measured from the FoF position of the halo, as 0.1R$_{200}$ < $r$ < R$_{200}$, to include only particles within the virial radius of the central while simultaneously avoiding the influence of the central galaxy's ISM. Note that this spatial criteria is well-matched to our satellite galaxy spatial selection (Section \ref{section: SatSampSelection}).

For this study, we are focused on the `cold' gas in the CGM, since the infalling gas feeding the central galaxy is mostly in this phase. We therefore use a temperature cut criterion, $T <$ 10$^{4.5}$ K to select most of our cold gas. However, in the ISM of any (central or satellite) galaxy, the two-phase pressurized model implemented in TNG gives rise to a temperature that is not strictly physical, but `effective'. This effective temperature of a star-forming gas cell is always higher in comparison to the temperature of the cold-phase in the subgrid model \citep{keres2012MNRAS.425.2027K,Sparre2016MNRAS.462.2418S}, as each star-forming subgrid gas cell also has a contribution from a hot subgrid phase. In the subgrid model, the unresolved cold phase has a temperature of 10$^3$ K and it dominates the mass-budget (in comparison to the unresolved hot phase), so in our analysis we override the `effective temperature' with a temperature of 10$^3$ K (for example, as adopted in \citet{ramesh2023MNRAS.518.5754R}). Thereafter, we define cold gas as all gas cells in TNG50 that satisfy the temperature cut criterion, $T <$ 10$^{4.5}$ K, as well as those cells having a non-zero instantaneous star formation rate.

As discussed in Sec. §\ref{section: Definitions}, each particle in a FoF halo has one of the following three associations-- Central, Satellites or the halo (Inner fuzz particles). 
We include the particles in all of these associations as the CGM of the central host, after which temperature constraints are applied to filter out cold gas cells. Using these assignations, we define two cases for the CGM and subsequently the cold CGM: with and without satellite gas. First, the \textit{`CGM+Satellites'} case includes all cold gas particles between 0.1R$_{200}< r <$ R$_{200}$, whether they are associated with the central galaxy, any satellites, or are part of the `Inner fuzz'. In the second case, which we refer to as \textit{`CGM-Satellites'}, we exclude all those particles that are 
spatially correlated with the satellites. We do so by applying a generous cut of 10.0 $\cdot$ \ R$_{0.5}$ (where R$_{0.5}$ is the stellar half mass radius of each satellite) for each satellite in our sample and exclude all cold gas cells within this radius\footnote{See Appendix §\ref{section: comp_dens} for a discussion on other values of radial cuts.}. Comparing these two cases allows us to determine how much cold CGM gas is close to satellite galaxies.

\section{Results}
\label{section: Results}

We have the mass in the cold gas phase in each halo with and without satellites (\textit{CGM+Satellites} and \textit{CGM-Satellites}, respectively), as defined in the previous section. We first obtain the spherical cold gas density profiles in order to investigate if the presence of satellites indeed has any impact on the extended cold gas profiles of their host halos. This is found by measuring the changes in cold gas mass with respect to three different properties of the satellite populations. To remind our readers, we exclude the ISM regions ($r <$ 0.1R$_{200}$) for every halo throughout our analysis.

\subsection{CGM cold gas density profiles}
\label{section: spherical}

Since satellites are gravitationally bound systems in themselves and lie within the potentials of their host galaxies, their presence is expected to have some impact on the internal mass distribution of their parent halos.
We might then expect differences in the radial spherical cold gas densities for the host galaxies depending both on the number of satellites bringing cold gas into the system, and on whether we include the gas still bound to satellites i.e. when comparing the \textit{CGM+Satellites} and \textit{CGM-Satellites} cases  (as described in §\ref{section:CGM}). 

\begin{figure*}
\centering
\includegraphics[width = \linewidth]{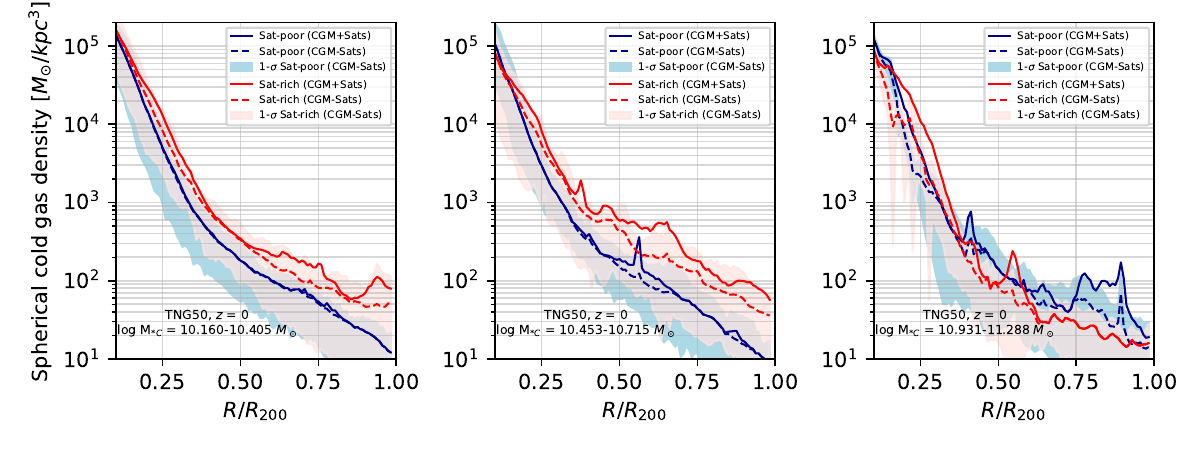}
\caption{\textit{Left-Right}: Comparison between the mean cold gas spherical densities (in M$_{\odot}$ kpc$^{-3}$) as a function of normalized radius for \textit{CGM+Satellites} (\textit{solid lines}) and \textit{CGM-Satellites} (\textit{dashed lines}), for TNG50 halos belonging to the low-mass (log M$_{\rm *C}$ = 10.160-10.405), middle-mass (log M$_{\rm *C}$ = 10.453-10.715) and high-mass bins (log M$_{\rm *C}$ = 10.931-11.288), for the 25th (\textit{blue}) and 75th (\textit{red}) quartiles. Regions within 0.1R$_{200}$ have been excluded to avoid ISM. Shaded blue (red) regions indicate 16-84th percentiles for the respective \textit{CGM-Satellites} cases. Note how differences start creeping between \textit{CGM+Satellites} and \textit{CGM-Satellites} mostly in regions at distances $\gtrsim$ 0.4R$_{200}$, directly indicating the influence satellites have on the cold gas profiles of galaxies.}
\label{fig:DensProfiles}
\end{figure*}

Accordingly, we plot the mean cold gas spherical densities for all our halos as a function of the normalized radius, from 0.1 to 1.0R$_{200}$, in Fig. \ref{fig:DensProfiles}. For the sake of clarity we do not plot the halos in the inter-quartile satellite number range in each mass bin. Satellite-poor (satellite-rich) halos are represented by blue (red) curves while the \textit{CGM+Satellites} (\textit{CGM-Satellites}) case is depicted by solid (dashed) curves. The shaded blue (red) regions indicate the 16-84th percentiles for the satellite-poor (satellite-rich) \textit{CGM-Satellites} cases. Each of the curves show the mean values of the population instead of medians is because we want to highlight the effect of satellites on the extended cold gas density profiles of the hosts and using medians suppresses their influence on the density profiles.

The changes between the similar colored solid and dashed lines show the cold gas associated with the satellite population in the halos. Across all our mass bins, we see smoother density profiles for the \textit{CGM-Satellites} cases (dashed), while the \textit{CGM+Satellites} cases (solid) show multiple density peaks, mostly in the regions 0.3 $< r <$ 1.0R$_{200}$. While there are indeed clear differences between the \textit{CGM+Satellites} and \textit{CGM-Satellites} curves, we have yet to find out how much cold gas, in terms of mass, is actually correlated with the satellite populations. We discuss this further in §\ref{section: Nsubs}, \ref{section: tms} and \ref{section: mms}.

On the whole, the cold gas densities drop sharply between 0.1 $< r <$ 0.35R$_{200}$ and more slowly between 0.35 $< r <$ 1.0R$_{200}$, thus giving rise to an ankle-like feature at $r \simeq$ 0.35R$_{200}$. This is reflected in the cumulative mass profiles (not shown) indicating that most of the mass resides in the central region. In both lower mass bins (log M$_{\rm *C}$ = 10.160-10.405 and log M$_{\rm *C}$ = 10.453-10.715), the satellite-rich halos generally have higher cold gas densities at $r >$ 0.25--0.35R$_{200}$ than their counterpart satellite-poor halos (although the shaded 1-$\sigma$ regions around the median values show significant overlap). The increase in mean density contrast between the satellite-poor and satellite-rich halos (and increasing offset of the 1$\sigma$ region) as a function of radius indicates that a satellite population tends to have more impact on the cold gas density in the outskirts of its associated halo. 


In the two lower-mass bins, the biggest difference between the \textit{CGM+Satellites} and \textit{CGM-Satellites} curves is also seen at larger radii. This aligns with our intuitive expectation that satellites tend to spend more time in the outskirts of their parent halos than they do when they approach the hosts at smaller radii. Satellites lying close to the halo center might also be completely stripped of gas and hence, may not be able to add more cold gas to the CGM \citep{putman2012}. Also as expected, the \textit{CGM+Satellites} and \textit{CGM-Satellites} curves are more offset for the satellite-rich samples than for the satellite-poor samples.

The high mass log M$_{\rm *C}$ = 10.931-11.288 sample has significantly fewer halos (just 21 halos) than the low mass (115 halos) and middle mass (61 halos) bin. Furthermore, the satellite-poor and satellite-rich subsets include only 5 halos each. Taking density averages over such a small sample can result in a strong impact by a few massive satellites. This is reflected in the few sharp intermittent density peaks in this bin profile, especially at large radii. Upon excluding the satellite contributions, these peaks dramatically smooth out. Unlike the other two bins, at large radii, the satellite-rich halos in this mass bin have slightly lower densities than the satellite-poor halos (even though the 1-$\sigma$ percentile shaded regions show a near-overlap and hardly any offset), irrespective of whether or not we include satellite contributions. However, we cannot determine if this shows different behavior at higher masses or if it is a false trend due to the low number statistics of this bin. Because of the small sample size of the log M$_{\rm *C}$ = 10.931-11.288 sample, throughout this paper, we will exercise caution when drawing any conclusions regarding halos in the high-mass bin. 

\subsection{CGM Cold Gas Mass}

The previous section (§\ref{section: spherical}) outlines the correlation between the cold gas density profiles of the host halos and the presence of satellites. The next step is to investigate the effect of various properties of the satellite population on the total cold gas mass (M$_{\rm cg}$) of the halos. 

\subsubsection{Number of satellites (N$_{\rm sats}$) and cold gas mass (M$_{\rm cg}$)}
\label{section: Nsubs}

\begin{figure*}
\centering
\includegraphics[width = \linewidth]{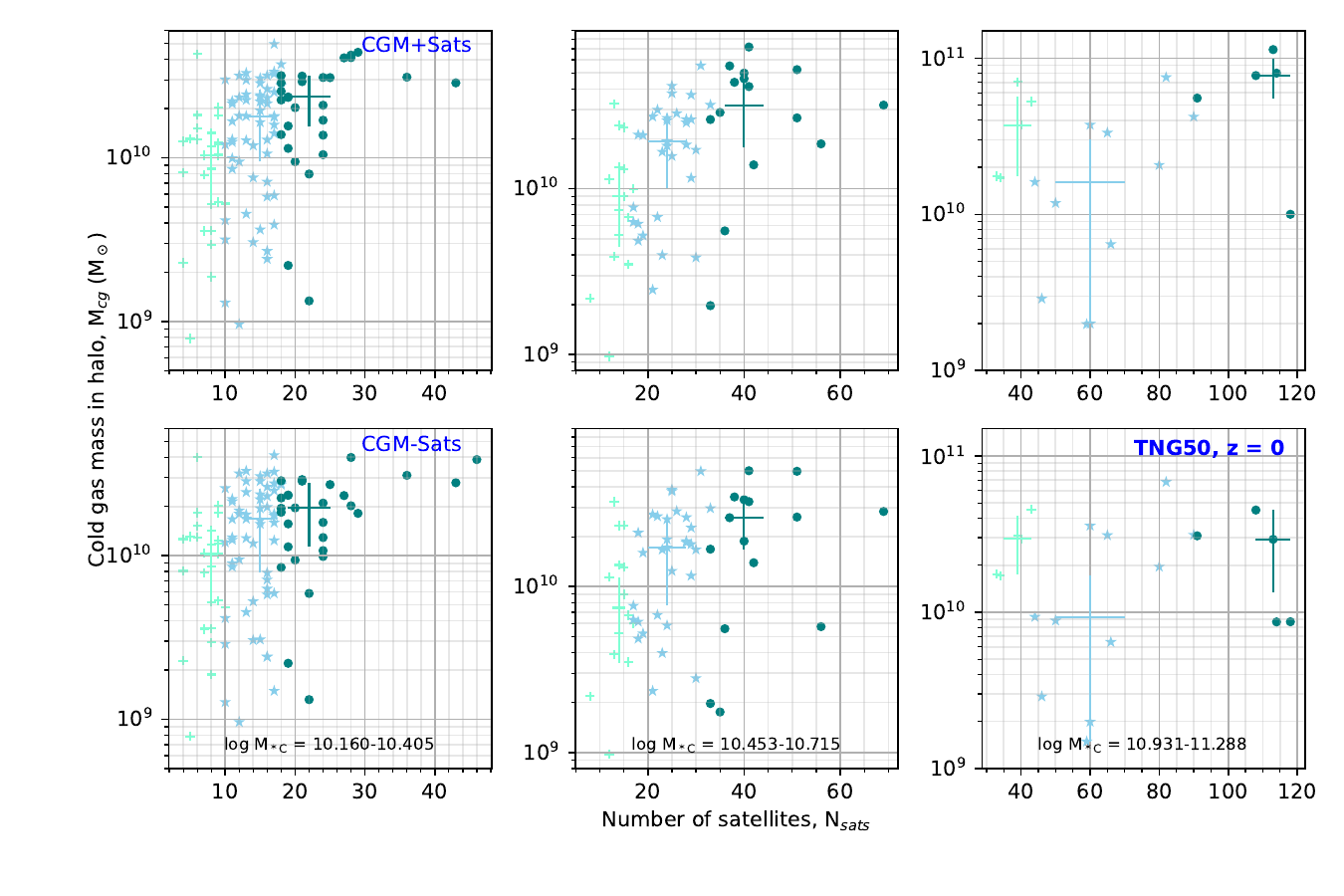}
\caption{Total cold gas mass, M$_{\rm cg}$, (in M$_{\odot}$) vs number of satellite galaxies per halo (N$_{\rm sats}$) for \textit{CGM+Satellites} (\textit{top panels}) and \textit{CGM-Satellites} (\textit{bottom panels}) for the log M$_{\rm *C}$ = 10.160-10.405 (\textit{left panels}), log M$_{\rm *C}$ = 10.453-10.715 (\textit{middle panels}) and log M$_{\rm *C}$ = 10.931-11.288 (\textit{right panels}) bin halos. Halos belonging to the 25th (\textit{plus symbols}), interquartile (\textit{star symbols}) and 75th (\textit{filled circles}) quartiles are shown for each bin. Respective median values with 1-$\sigma$ ranges (\textit{colored cross symbols}) are shown for each quartile. In all three mass bins, when including gas associated with satellite galaxies (top panels), there is more M$_{\rm cg}$ in satellite-rich systems than in satellite-poor systems.}
\label{fig:NumSats}
\end{figure*}

We begin with the number of satellites per halo (N$_{\rm sats}$) and plot these against the corresponding total CGM cold gas mass (M$_{\rm cg}$) values in Fig. \ref{fig:NumSats}. We compute the M$_{\rm cg}$ values for halos in the \textit{CGM+Satellites} (top panels) and \textit{CGM-Satellites} (bottom panels) cases and plot them (
satellite-poor halos are cyan crosses, the interquartile halos are blue stars and the satellite-rich halos are dark circles)) against N$_{\rm sats}$. Note that the $x$- and $y$-axes differ across the mass bins. Respective medians (${\rm \bar N_{\rm sats}}$, ${\rm \bar M_{\rm cg}}$) for halos in each of the satellite-rich and satellite-poor quartiles as well as the interquartile region along with their 1-$\sigma$ ranges are also overplotted as colored large cross symbols. 


As we move from the lowest mass galaxies (left panels) to the highest mass galaxies (right panels) we see that the M$_{\rm cg}$ values tend to increase, but the scatter is large enough for all bins to have significant overlap in M$_{\rm cg}$.  If we focus on the upper \textit{CGM+Satellites} panels, in all mass bins the satellite-rich galaxies have larger median M$_{\rm cg}$ than the satellite-poor galaxies. For both the log M$_{\rm *C}$ = 10.160-10.405 (left panel) and log M$_{\rm *C}$ = 10.453-10.715 (middle panel) mass bins, there is no overlap between the satellite-poor and satellite-rich 1-$\sigma$ ranges and only a slight overlap in the highest mass bin (right panel). Indeed, for both the lower mass bins, ${\rm \bar M_{\rm cg}}$ increases nearly linearly as we go from satellite-poor halos to satellite-rich halos. This trend does not occur in the log M$_{\rm *C}$ = 10.931-11.288 (right panels) high mass bin. However, again we stress that it is not possible to robustly conclude much owing to the small sample size for this bin.


Next, we compare the amount of cold gas when including satellite gas (\textit{CGM+Satellites}, top panels) to that after excluding satellite gas (\textit{CGM-Satellites}, bottom panels). For log M$_{\rm *C}$ = 10.160-10.405 bin, the satellite-poor M$_{\rm cg}$ 
values vary very little, irrespective of whether or not we include satellite contributions; in other words, for these halos, most of the cold gas resides outside of the satellites. 
This trend for satellite-poor galaxies is also corroborated by the near overlap between satellite-poor density profiles for \textit{CGM+Satellites} and \textit{CGM-Satellites} cases in the leftmost panel in Fig. \ref{fig:DensProfiles}. In all other galaxy subsamples, the ${\rm \bar M_{\rm cg}}$ value decreases when comparing the \textit{CGM+Satellites} to the \textit{CGM-Satellites} values. For all three mass bins, the difference in ${\rm \bar M_{\rm cg}}$ values between the top and bottom panels increases slowly as we go from satellite-poor to satellite-rich halos. In fact, in the satellite-rich quartile of the log M$_{\rm *C}$ = 10.931-11.288 bin the decrease in ${\rm \bar M_{\rm cg}}$ is so large that their is no ${\rm \bar M_{\rm cg}}$ difference between the satellite-poor and satellite-rich samples. Such a difference is indicative of the existence of some cold gas associated with the satellite populations, and more cold gas associated satellites in larger populations and higher mass halos.

To summarize, we see an overall correlation between the number of satellites and the cold gas mass in the halo. However, the difference between the ${\rm \bar M_{\rm cg}}$ of the satellite-poor and satellite-rich populations is more likely to be significant in lower mass halos and when including cold gas near satellites (\textit{CGM+Satellites}). 

\subsubsection{Total mass in satellites (M$_{\rm tms}$) and cold gas mass (M$_{\rm cg}$)}
\label{section: tms}
\begin{figure*}
\centering
\includegraphics[width = \linewidth]{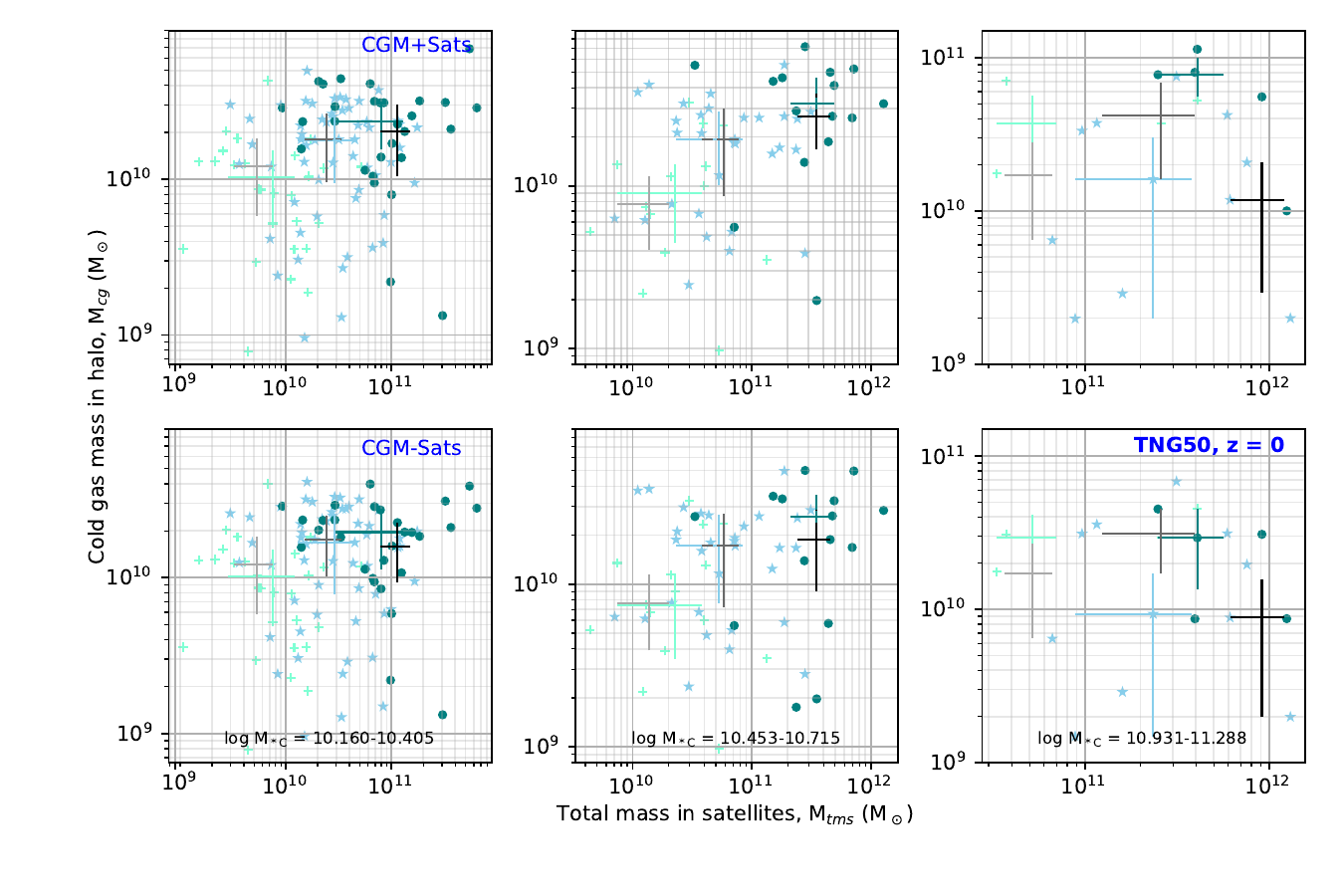}
\caption{Total cold gas mass per halo, M$_{\rm cg}$, (in M$_{\odot}$) vs Total mass in all satellite galaxies per halo, M$_{\rm tms}$ (in M$_{\odot}$) for \textit{CGM+Satellites} (\textit{top panels}) and \textit{CGM-Satellites} (\textit{bottom panels}) for the log M$_{\rm *C}$ = 10.160-10.405 (left panels), log M$_{\rm *C}$ = 10.453-10.715 (middle panels) and log M$_{\rm *C}$ = 10.931-11.288 (right panels) bin halos. As in Fig. \ref{fig:NumSats}, number of satellites-based medians (${\rm \bar N_{\rm sats}}$) are represented by colored cross symbols. M$_{\rm tms}$-based median values (${\rm \bar M_{\rm tms}}$) are shown as light grey, grey and black crosses respectively.}
\label{fig:MassSats}
\end{figure*}

The total mass in satellites should correlate strongly with the number of satellites, but there will be scatter in this relation. More massive satellite galaxies could harbor larger reserves of cold gas, potentially contributing more towards the total cold gas budget of its halo. Therefore, M$_{\rm cg}$ may correlate more strongly with the total mass of satellites than the total number of satellites. To test this, we plot the total mass (sum of M$_*$, M$_{\rm gas}$ and M$_{\rm h}$) in satellites (M$_{\rm tms}$) for each halo against the corresponding M$_{\rm cg}$ values in Fig. \ref{fig:MassSats}. 

In this figure, colored points and symbols denote the N$_{\rm sats}$ position, just as in Figure \ref{fig:NumSats}. Examining the distribution of points in Fig. \ref{fig:MassSats} we see that N$_{\rm sats}$ does not perfectly correlate with M$_{\rm tms}$, as expected. In other words, some halos having a low number of satellites actually have more total satellite mass than some halos with a higher number of satellites. This imperfect correlation between N$_{\rm sats}$ and M$_{\rm tms}$ results in a mixing of the points with different symbols along the x-axis (unlike in Fig. \ref{fig:NumSats} where there was a clear x-axis range for each quartile). 
Therefore, to compare the N$_{\rm sats}$-M$_{\rm cg}$ correlation to the M$_{\rm tms}$-M$_{\rm cg}$ correlation we plot two sets of median values here. As before, using the color scheme in Fig. \ref{fig:NumSats}, we plot the medians when ranked according to the N$_{\rm sats}$ (${\rm \bar N_{\rm sats}}$; green/blue crosses).
Note that by construction the $y$-values of each of these crosses is the same as in Fig. \ref{fig:NumSats}. Next, we plot the medians of a new set of quartile ranges computed after ranking the M$_{\rm tms}$ values in ascending order (${\rm \bar M_{\rm tms}}$; grey/black crosses). This gives us a clear picture of how the cold gas mass evolves across quartiles within a bin once we arrange the halos according to their M$_{\rm tms}$ values. In this subsection we will base our interpretations on ${\rm \bar M_{\rm tms}}$ rather than ${\rm \bar N_{\rm sats}}$.


For the low mass log M$_{\rm *C}$ = 10.160-10.405 bin halos, the upper panel cold gas mass ${\rm \bar M_{\rm tms}}$ value increases very gradually, and the overlapping 1-$\sigma$ ranges of the low-M$_{\rm tms}$ quartiles and high-M$_{\rm tms}$ quartile indicate the trend is not significant. The relatively steeper rise seen in case of log M$_{\rm *C}$ = 10.453-10.715 bin halos does show a significant difference in the cold gas mass in low-M$_{\rm tms}$ versus high-M$_{\rm tms}$ halos. The median cold gas mass in the log M$_{\rm *C}$ = 10.931-11.288 halos increases initially but drops as we transition into the satellite-rich quartile, and all the 1-$\sigma$ ranges overlap. However, as stated earlier, a larger sample size is needed to test the veracity of such a trend. 
As above, when removing the cold gas near satellites, the lower \textit{CGM-Sats} panels, all trends are weakened.

Thus, while we see an overall positive trend in M$_{\rm cg}$ when we consider M$_{\rm tms}$ of satellites, it is only significant for the log M$_{\rm *C}$ = 10.453-10.715 mass bin when including the gas near satellites (\textit{CGM+Satellites}).



\subsubsection{Stellar mass of the most massive satellite (M$_{\rm *mms}$) and cold gas mass (M$_{\rm cg}$)}
\label{section: mms}

\begin{figure*}
\centering
\includegraphics[width = \linewidth]{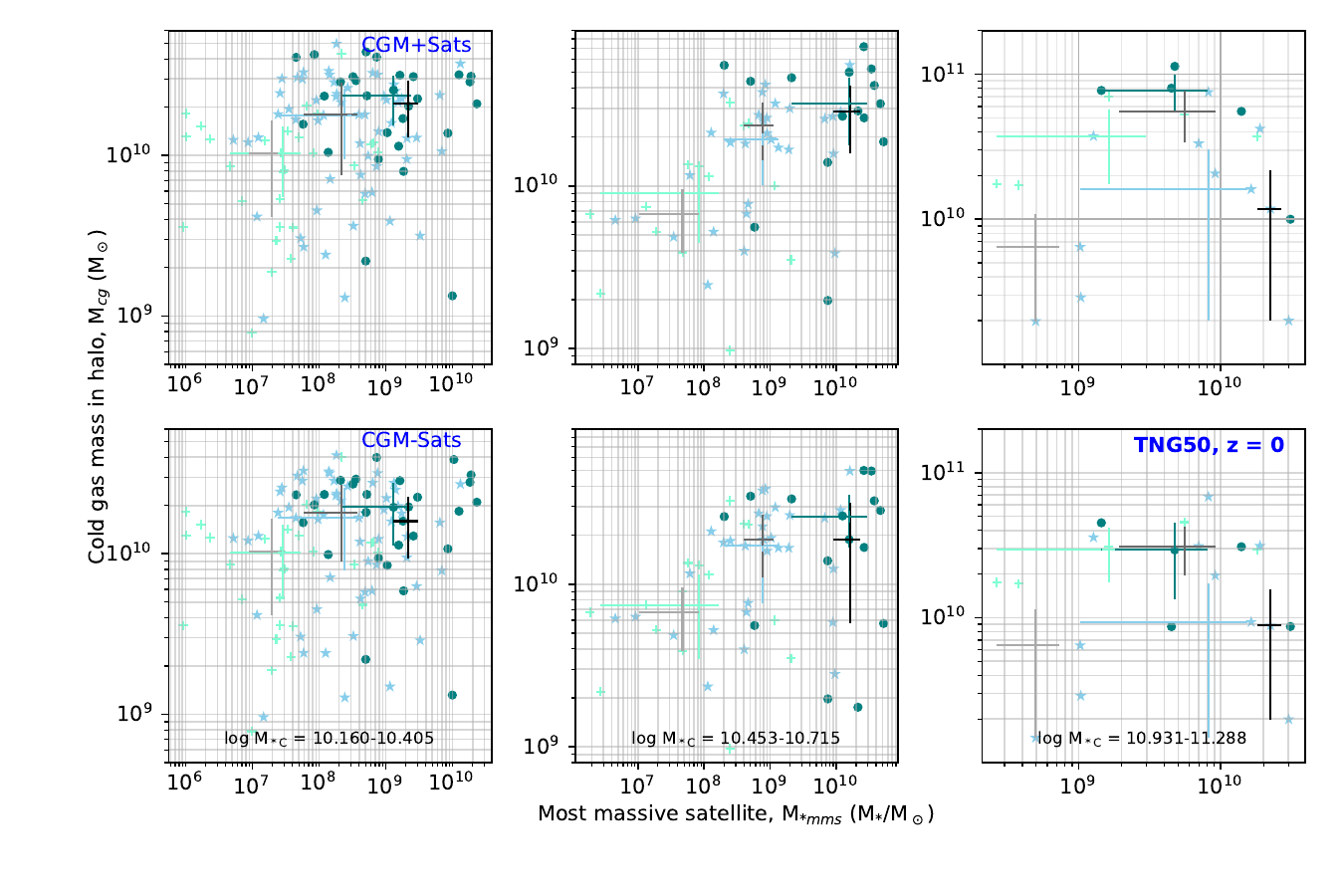}
\caption{Total cold gas mass per halo, M$_{\rm cg}$, (in M$_{\odot}$) vs the stellar mass of the most massive satellite galaxy in each halo, M$_{\rm *mms}$ (in M$_{\odot}$) for \textit{CGM+Satellites} (\textit{top panels}) and \textit{CGM-Satellites} (\textit{bottom panels}) for the log M$_{\rm *C}$ = 10.160-10.405 (left panels), log M$_{\rm *C}$ = 10.453-10.715 (middle panels) and log M$_{\rm *C}$ = 10.931-11.288 (right panels) bin halos. As in Fig. \ref{fig:MassSats}, respective quartiles and 1-$\sigma$ ranges are marked.}
\label{fig:MostMassiveSat}
\end{figure*}

So far we have considered the entire satellite population in each halo, which is likely to be largely composed of low-mass satellites that may not contribute a significant amount of gas to the CGM or be observable with current facilities. High-mass satellites with correspondingly large stellar and gas mass content could deposit a substantial fraction of the total cold gas into the CGM of their host halos \citep{Deason_2016,fattahi10.1093/mnras/staa2221} and may also be more readily observable than the less massive satellites. Therefore, we choose the stellar mass of the most massive satellite per halo (M$_{\rm *mms}$) as our third property and plot it against M$_{\rm cg}$ in Fig. \ref{fig:MostMassiveSat}. 


Just like in Fig. \ref{fig:MassSats}, we compute the medians after ranking the halos according to M$_{\rm *mms}$ (${\rm \bar M_{\rm *mms}}$; greyscale crosses), in addition to our original medians using ${\rm \bar N_{\rm sats}}$. We will again base our interpretations on ${\rm \bar M_{\rm *mms}}$ rather than ${\rm \bar N_{\rm sats}}$. ${\rm \bar M_{\rm *mms}}$ show a slightly steeper increase in M$_{\rm cg}$ for the low mass log M$_{\rm *C}$ = 10.160-10.405 bin halos as compared to that in Fig. \ref{fig:MassSats}, although the 1-$\sigma$ ranges for the satellite-rich and satellite-poor still overlap. 
The ${\rm \bar M_{\rm *mms}}$ values show a larger increase as compared to that for Figs. \ref{fig:NumSats} and \ref{fig:MassSats} for the middle mass log M$_{\rm *C}$ = 10.453-10.715 halos. We again see significant scatter in the high-mass bin halos but refrain from drawing any conclusions for this small sample.

In summary, M$_{\rm *mms}$ brings a consistent positive change in M$_{\rm cg}$ for the low and middle mass halos when including gas near satellites (CGM+Satellites). 



\section{Discussion}
\label{section: Discussion}


In this work we have analyzed the extended cold gas distributions for low-, middle- and high-mass TNG50 galaxies from the perspective of three different satellite properties, N$_{\rm sats}$, M$_{\rm tms}$ and M$_{\rm *mms}$. In this section, we present our main reflections from these results. 
We expected that the amount of cold gas contained in a halo is most closely tied to the central galaxy's mass before having any dependence on the global satellite properties, and, indeed, this agrees with our findings (see §\ref{section:final}). 
We will first present our interpretations regarding which of the considered satellite properties shows the strongest correlation with the cold gas mass (§\ref{section: matters}). To compare with observations beyond the most massive satellite, we count the number of massive satellites per halo and their correlation with the cold gas mass (§\ref{section: number}). We follow this up with a short discussion analyzing the cold gas mass associated with the most massive satellite and what that tells us about the cold gas stripping trends (§\ref{section: indications}). Finally, we include a brief summary on other cold gas sources in galaxies as well as the relevance of this study in the context of the Local Group (§\ref{section: supply} \& §\ref{section: implications}).

\subsection{Satellite population: What matters the most?}
\label{section: matters}

It is clear from Figs. \ref{fig:NumSats}, \ref{fig:MassSats} and \ref{fig:MostMassiveSat} that N$_{\rm sats}$ correlates positively with M$_{\rm cg}$ across the entire sample. We expect each of these satellite population properties to correlate with one another. A higher N$_{\rm sats}$ should, in general, translate to a higher value of M$_{\rm tms}$. In addition, because the stellar mass function of galaxies has a negative slope, the larger the number of satellites in a halo, the more likely we are to find a massive satellite among them. In other words, a higher N$_{\rm sats}$ should also translate to a higher M$_{\rm *mms}$. In the absence of scatter in the stellar mass function for satellites we would have expected to find the exact same relationship between satellite-poor and satellite-rich quartiles, irrespective of the satellite property we used. The existence of the general trends between M$_{\rm cg}$ and the satellite properties is, therefore, expected. However, finding out which of these trends is the strongest can help guide observers on how they should select gas-rich CGM targets.

From the upper panels in Figs. \ref{fig:NumSats}, \ref{fig:MassSats} and \ref{fig:MostMassiveSat}, N$_{\rm sats}$ causes a maximal increase in M$_{\rm cg}$ for the low-mass halos. This is also the only correlation for which the median plus 1-$\sigma$ regions of the satellite-poor and satellite-rich regions do not overlap. 
For the middle mass bin halos, even though all three variables cause a significant increase in M$_{\rm cg}$ between the satellite-poor and satellite-rich samples using the 1-$\sigma$ ranges, the M$_{\rm *mms}$ shows the most separation. For the massive halos,  N$_{\rm sats}$ and M$_{\rm *mms}$ show a positive correlation from the satellite-poor to satellite-rich quartiles while M$_{\rm tms}$ shows a negative slope (with all three properties showing an inflection via the inter-quartile halos). From the minimal overlap of 1-$\sigma$ medians, it does seem that, when considering the full mass range, N$_{\rm sats}$ correlates best with M$_{\rm cg}$. Large sample sizes would be required, however, to more robustly test this. 

Thus, N$_{\rm sats}$ correlates the most with cold gas mass for low-mass and high-mass halos (with the caveat that our high mass sample is very small), while M$_{\rm *mms}$ is the most strongly correlated property for middle-mass halos. 
M$_{\rm *mms}$ is also the most straightforward observable for Milky Way and sub-Milky Way mass halos, allowing this correlation to be tested. N$_{\rm sats}$, however, is not an entirely realistic observable since it is impossible to detect the complete satellite population including galaxies down to the lowest end of stellar mass values\footnote{Typical observational stellar mass detection limits at low-$z$ for satellites is M$_{*} >$ 10$^6$  M$_{\odot}$ \citep{martin10.1111/j.1365-2966.2007.12055.x, Kirby_2013}.} (this difficulty increases manifold for non-Local Group galaxies). In Sec. §\ref{section: number} we now determine how the massive satellites' population correlates with CGM cold gas mass and probe whether these can instead be used as a proxy for N$_{\rm sats}$.

\subsection{Number of massive satellites associated with a halo}
\label{section: number}
In addition to being more easily detected, substantially massive satellites are more likely to bring a significant amount of cold gas into a host CGM \citep{roy2023MNRAS.tmp.3042R}. Several studies adopt a mass-cut definition for demarcating low-mass from high-mass satellites. Usually, satellites having M$_{*} <$ 10$^8$  M$_{\odot}$ are considered as \textit{low-mass} satellites, while those having M$_{*} >$ 10$^8$  M$_{\odot}$ are considered as \textit{massive} satellites \citep{McConnachie_2012, Weisz_2015}. Using this definition, we now look at the trend between the number of massive satellites and M$_{\rm cg}$.

\begin{figure*}[h!]
\centering
\includegraphics[width = \linewidth]{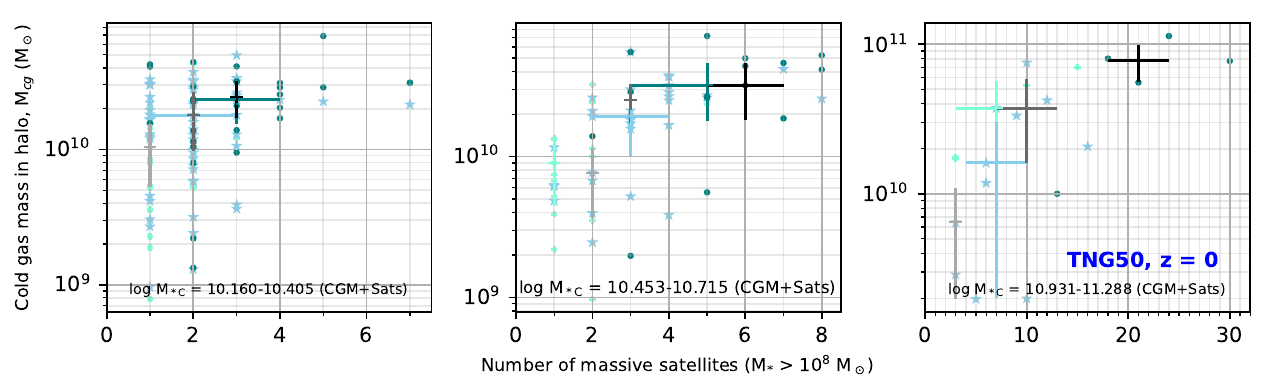}
\caption{Total cold gas mass per halo, M$_{\rm cg}$ (in M$_{\odot}$) vs the number of massive (M$_{*} >$ 10$^8$ M$_{\odot}$) satellites in a halo. Left, middle and right panels show log M$_{\rm *C}$ = 10.160-10.405, log M$_{\rm *C}$ = 10.453-10.715 and log M$_{\rm *C}$ = 10.931-11.288 bin halos. As in Fig. \ref{fig:MassSats}, respective quartiles and 1-$\sigma$ ranges are marked. Note that, in the highest mass bin, the median satellite-rich quartile, ranked according to N$_{\rm sats}$ (dark green cross) exactly overlaps with the corresponding median ranked according to the number of massive satellites (black cross).}
\label{fig:massive}
\end{figure*}

In Fig. \ref{fig:massive}, we plot the number of massive satellites versus the cold gas mass of their host galaxies. We count the number of satellites which have M$_{*} >$ 10$^8$  M$_{\odot}$ and compute the total amount of cold gas in that halo. Like in the previous plots, we show medians when ranking according to N$_{\rm sats}$ (blue-green cross symbols). Because the galaxies in the low and middle mass bins have a relatively small range of massive satellites, it is not possible to represent quartiles using the number of massive satellites in a fair manner. Therefore, for the low mass bin we show the median and 1-$\sigma$ of galaxies with 1 massive satellite (44 galaxies), 2 massive satellites (39 galaxies), and 3 or more massive satellites (32 galaxies). In the middle mass bin we show the median of galaxies with 1-2 massive satellites (26 galaxies), 3-4 massive satellites (19 galaxies) and 5 or more massive satellites (16 galaxies). Since the highest mass bin halos have a larger range in their number of massive satellites, we are able to show quartiles simply by ranking according to the number of massive satellites. These medians are represented by greyscale crosses.

The median number of massive satellites varies from 1 to 3, 2 to 6 and 3 to 21 from the satellite-poor to the satellite-rich quartiles for the low-mass, middle-mass and high-mass halos respectively. From Fig. \ref{fig:NumSats} we see that the median total number of satellites for the three mass bins ranged from 8 to 22, 14 to 40 and 40 to 113 satellites. Thus, we see a factor of $\sim$3-4 increase in massive satellites, similar to what we see for the total number of satellites. As we would expect, the most massive halos host the largest pool of massive satellites, as well as the largest total number of satellites. 


Although the total number of satellites down to low masses is difficult to observationally measure, a census of more massive satellites, M$_{*} >$ 10$^8$  M$_{\odot}$, is much more reasonable.  Therefore, it is heartening for future surveys that we find the correlation between the cold CGM mass and the number of massive, more easily observed, satellites to be nearly identical to the correlation between the cold CGM mass and the total number of satellites. Indeed, although the sample size is small, we note that in the highest mass bin the correlation between the number of massive satellites and the cold gas mass in the CGM is stronger than any other satellite population feature we have studied here.
 However, even though the 1$\sigma$ ranges do not overlap in any mass bin, we stress that there is a large scatter in this relation, meaning that a large number of halos would need to be observed ranging from few to many massive satellites 
before any trend becomes statistically significant. 

Our results do not prove, but only mildly suggest, that massive satellites may be important in feeding cold gas into the CGM. This agrees with other recent simulation studies as well. \citet{roy2023MNRAS.tmp.3042R} ran a suite of idealized MW-like halos varying the satellite populations and found that more massive satellites retained their gas longer, and once stripped, the gas tended to remain cold in the CGM. \citet{ramesh2023MNRAS.518.5754R} find that in TNG50 MW-like halos, cold gas clouds tend to lie near satellites. Based on our results and \citet{roy2023MNRAS.tmp.3042R}, we would expect the cold clouds to lie near more massive satellites, whose stripped gas can survive for significant amounts of time in the CGM (more than 1 Gyr).

\subsection{Indications about cold gas stripping}
\label{section: indications}

\begin{figure*}[h!]
\centering
\includegraphics[width = \linewidth]{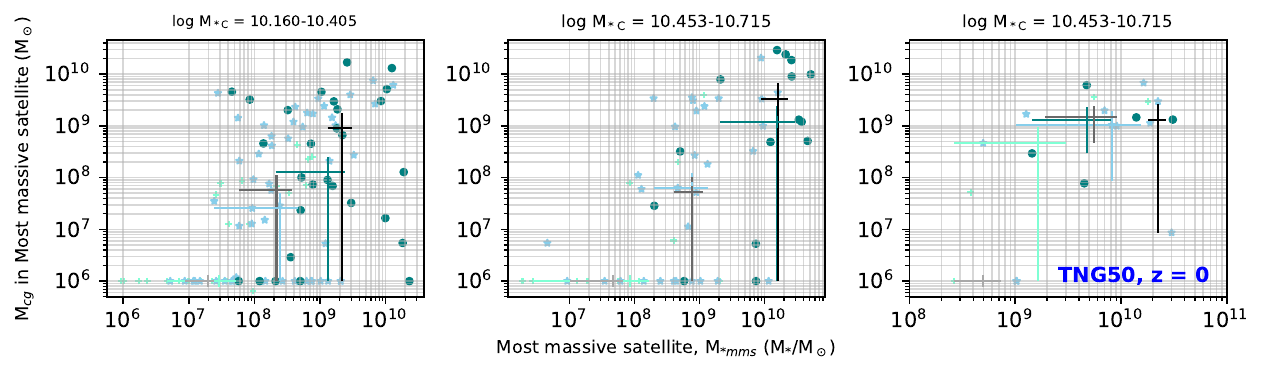}
\caption{Total cold gas mass, M$_{\rm cg}$, (in M$_{\odot}$) in the most massive satellite vs the stellar mass of the most massive satellite galaxy, M$_{\rm *mms}$. Left, middle and right panels show log M$_{\rm *C}$ = 10.160-10.405, log M$_{\rm *C}$ = 10.453-10.715 and log M$_{\rm *C}$ = 10.931-11.288 bin halos. As in Fig. \ref{fig:MassSats}, respective quartiles and 1-$\sigma$ ranges are marked. Since some most massive
satellites are totally devoid of any cold gas, we explicitly set their y-axis values to $10^6$ M$_{\odot}$. The amount of cold gas contained within each most massive satellite gradually increases as we move from low mass to high mass halos.}
\label{fig:cgmms}
\end{figure*}

While we do not explicitly track the cold gas mass evolution in our study, it is nevertheless possible to gain some insights about the cold gas stripping by calculating how much cold gas mass is associated with the ISM and the CGM of the most massive satellite in each halo in the $z$$=$0 snapshot. We remind the readers that we define the bounds of a satellite at $r \sim$  10.0 $\cdot R_{0.5}$ and assume that the regions within this radius constitute the satellite ISM and CGM while the rest are dissociated from the satellite. 
This leads us to Fig. \ref{fig:cgmms}, which shows the cold gas mass spatially associated with the most massive satellite as a function of its stellar mass. 

We explicitly set the M$_{\rm cg}$ values of satellites without any associated cold gas to 10$^6$ M$_{\odot}$, purely for the purpose of facilitating median calculations. As before, we focus on the greyscale medians. The overall massive satellite ${\rm \bar M_{\rm cg}}$ increases linearly with M$_{\rm *mms}$ for the low- and middle-mass halos, while the median values plateau for high-mass halos. However, since the subsample size of the satellite-rich halos is quite low (5 halos), we do not consider this a robust trend. It is interesting, however, to note that the amount of cold gas retained in the satellites associated with the massive halos is roughly similar to that retained in the satellites associated with the low-mass halos. Directly comparing the satellite-rich halos in both bins, we see that even though the most massive satellite in the high-mass bin is an order of magnitude more massive as compared to the low-mass one, their cold gas mass is comparable. This may indicate that massive halos strip gas away from their satellites more efficiently than less massive halos (though further analysis and a more robust sample size is required to confirm this conclusion).

Satellite-poor, low-mass host galaxies tend to have most massive satellites with lower mass values. However, not all satellites have cold gas associated with them. In fact, the satellite stellar mass at which we start seeing cold gas associated with a satellite increases as we increase the host mass. The low mass bin has a few satellites with gas at M$_{\rm *} \gtrsim$ 3 $\times$ $10^7$ M$_{\odot}$, while we do not see any cold gas associated with satellites associated with hosts in middle mass and high mass bins until they reach M$_{\rm *} \gtrsim$ 1 $\times$ $10^8$ M$_{\odot}$ and M$_{\rm *} \gtrsim$ 3 $\times$ $10^8$ M$_{\odot}$ respectively. Isolated galaxy simulation studies (see Fig. 4 and Table 1 in \citet{roy2023MNRAS.tmp.3042R}) echo our results and find that less massive satellites (M$_{\rm *}$ $\lesssim$ $10^7$ M$_{\odot}$) were stripped of their cold gas within a Gyr while a more massive satellite (M$_{\rm *} \gtrsim$ $10^8$ M$_{\odot}$) is able to retain its cold gas reservoir for a long time. Although the satellites in our sample extend to lower masses than in \citet{stevens2019MNRAS.483.5334S, donnari2021MNRAS.500.4004D}, our work also agrees with these larger-scale studies of the TNG simulation suite that find that satellites in more massive halos tend to have higher quenched fractions and lower gas fractions, and that at the same halo mass, lower mass satellites have higher quenched fractions and lower gas fractions.

\subsection{The supply chain of cold gas in galaxies}
\label{section: supply}

\begin{figure*}
\centering
\includegraphics[width = \linewidth]{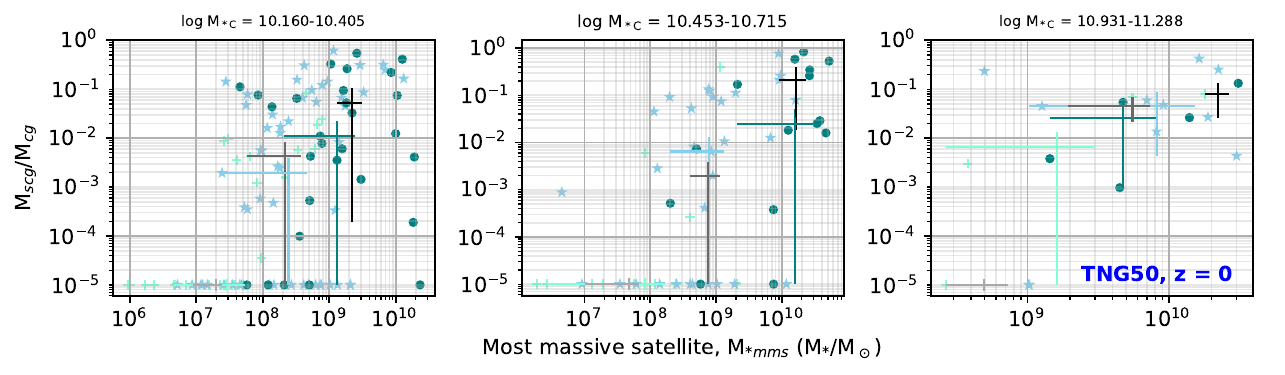}
\caption{Total cold gas mass within $R_{10}$ of all satellites (M$_{\rm scg}$) upon the total halo cold gas mass (M$_{\rm cg}$) vs Stellar mass of the most massive satellite, M$_{\rm *mms}$ (in M$_{\odot}$). Left, middle and right panels show log M$_{\rm *C}$ = 10.160-10.405, log M$_{\rm *C}$ = 10.453-10.715 and log M$_{\rm *C}$ = 10.931-11.288 bin halos. As in earlier figures, respective quartiles and 1-$\sigma$ ranges are marked. Since some satellite populations are totally devoid of any cold gas within their $R_{10}$ regions, we set their y-axis values to 10$^{-5}$. This figure helps us directly quantify the ratio of the total halo cold gas mass coming from the $R_{10}$ regions of the satellite population. While the median contributions from satellites towards the host cold gas mass are less than 10\%, some individual cases show as much as $\sim$ 70-80\% of the halo cold gas associated with their corresponding satellite populations.
}
\label{fig:satgas}
\end{figure*}

In this paper, we have found correlations between the halo cold gas mass  and their satellite populations. Even though there exists a difference when comparing the M$_{\rm cg}$ after including satellites (\textit{CGM+Satellites}) and excluding satellites (\textit{CGM-Satellites}), this difference is often small. Thus, it is clear that gas spatially associated with satellite galaxies is not likely to be the most common major contributors towards the halo cold gas at low-$z$. To further quantify the fraction of cold gas within satellite galaxies, in Figure \ref{fig:satgas}, we plot the ratio of total cold gas mass lying within $R_{10}$ of any the satellites ({M$_{scg}$; total cold gas mass in all satellites) to the total halo cold gas mass ({M$_{\rm cg}$) in the \textit{CGM+Satellites} case, against the stellar mass of the most massive satellite (M$_{\rm *mms}$)}}. We see that, in most cases, less than 10\% of the host CGM cold gas mass is associated with satellites. In some cases, the fraction of cold gas can be tens of percent, with few galaxies having more than 60\% of its cold CGM gas associated with satellites. Unsurprisingly, given Fig. \ref{fig:cgmms}, it seems that there is more cold gas mass associated with satellites as we increase the mass of the most massive satellite\footnote{Barring again, the small number of high-mass halos that do not show a very clear increasing trend.}, but there is significant scatter across all the halos. We note that just because cold gas is far from a satellite galaxy does not mean it could not have originated in a satellite \citep{roy2023MNRAS.tmp.3042R}. This, however, still suggests that gas coming directly from satellites is often a small component of the cold CGM reservoir.

Thus, below we discuss the other sources of cold gas in the CGM, moving from the outer CGM to the central galaxy.  After introducing each source we determine if and how they relate to the satellite component of CGM cold gas.  


 
\subsubsection{Filamentary accretion}
\label{section: filament}

Filamentary accretion, in the form of metal-poor gas, from the IGM is one of the most conventional channels through which a galaxy acquires its supply of cold gas \citep{nelson2013MNRAS.429.3353N}. For massive galaxies, most of this incoming gas gets shock heated to virial temperatures as it passes through the galactic gravitational potential well-- so-called hot-mode accretion \citep{white1978MNRAS.183..341W, white1991ApJ...379...52W, birnboim2003MNRAS.345..349B, walters10.1093/mnras/stab840}. However, for the halos considered here, the halo potential is only able to heat up the accreting gas to intermediate temperatures and some gas may come in as unshocked filaments, aka the cold mode accretion \citep{keres2005MNRAS.363....2K,dekel2009Natur.457..451D,nelson2016MNRAS.460.2881N,mandelker2020MNRAS.494.2641M}. In addition, the heated-up gas, later, cools down and rains upon the ISM, eventually contributing to the star-formation process \citep{marinacci2010MNRAS.404.1464M}. Indications from group-scale halos suggests that some of this cooling may also be caused by satellite galaxies \citep{saeedzadeh2023MNRAS.525.5677S}, although, it remains to be seen whether this holds true in case of galaxies in our mass ranges (in the small idealized suite in \citet{roy2023MNRAS.tmp.3042R} the satellites did not induce cooling). Purely from a density field perspective, a link between the satellite population and filamentary accretion surrounding the host is not just expected but also observed \citep{tempel2015MNRAS.450.2727T, wang2020ApJ...900..129W}. 

While a comprehensive study focusing on disentangling the connection between the specifics of filamentary accretion and satellite populations is beyond the scope of this paper, we expect that more satellites would generally translate to the presence of stronger filamentary accretion around a galaxy. We posit that extra cold gas from accretion along filaments and satellite-induced cooling would likely be found in the outskirts of a halo, where the main differences in the cold gas densities between the satellite-rich and satellite-poor samples are seen in Figure \ref{fig:DensProfiles}.  

\subsubsection{Galactic fountain flows}
\label{section: fountain}

Gas flows between the ISM and halo regions, aka galactic fountain flows, is another potentially important pathway for cold gas to enter the CGM \citep{fraternali10.1111/j.1365-2966.2005.09816.x, fraternali2008}. Hot, fast winds driven by supernova feedback are propelled out to CGM distances. On their way out to CGM, the baryons in these winds decelerate and lose energy, eventually cooling down and raining back on to the galaxy. The growth of cool clouds is accelerated via the interaction of feedback-driven winds with the surrounding hot, ambient medium present outside the ISM \citep{alcazar2017MNRAS.470.4698A, suresh10.1093/mnras/sty3402}. We recall that the SFR values for the satellite-poor and satellite-rich galaxies within each bin are generally similar (see Figs. \ref{fig:LMBScatter}, \ref{fig:MMBScatter} and \ref{fig:HMBScatter}). This would, therefore, roughly translate to a SNe feedback of similar magnitudes. Thus, in the context of the above discussion, this feedback could end up affecting the central CGM spherical cold gas densities in a similar manner \citep{Fielding_2020}. Indeed, the overall density profiles for \textit{CGM-Satellites} case for every bin show little difference (see the dashed red and blue lines in Fig. \ref{fig:DensProfiles}) within 0.15R$_{200}$. This supports our argument that the magnitude of SNe feedback affecting the central CGM cold gas densities is similar across all galaxies. Recall from Figure \ref{fig:DensProfiles} that the cold gas density is much higher in the inner regions of the halo. If this gas is mainly from galactic fountain flows, these may be one of the major sources of cold CGM gas. However, because these fountain flows generally reside in the inner CGM, we posit that there is little evidence that SNe feedback from the host central galaxy should affect the satellite cold gas component in the CGM.

\subsubsection{SMBH feedback}
\label{section: SMBH}

Finally, we discuss SMBH feedback, which can have an impact on the CGM outer to larger radii than fountian flows. For details of the AGN accretion and feedback implementation in the TNG suite, we point the reader to \citet{weinberger2018MNRAS.479.4056W, pillepich2021MNRAS.508.4667P}. Here we focus on the effects of AGN feedback in TNG on the host CGM. Radio-mode galactic feedback, arising from accreting SMBHs in the low-$z$ universe, can alter the CGM phase distributions to some extent. Resultant AGN winds drive large radio bubbles at kpc distances, which could impact future IGM accretion and transform some of the cold-dense gas phase to a warmer, more diffuse phase \citep{zinger10.1093/mnras/staa2607}. In their TNG50 study of MW-like galaxies, \citet{ramesh2023MNRAS.518.5754R} found evidence for SMBH-driven fast outflows as well as the heating of halo gas to super-virial temperatures-- ultimately regulating the net balance of inflows and outflows. Focusing on the direct impact of AGN feedback on cold gas, however, \citet{ma2022ApJ...941..205M} find that kinetic feedback in TNG is weak and only moves cold gas in the inner disk to slightly larger galactocentric distances. 

\citet{davies2020MNRAS.491.4462D} probed the connection between CGM mass fraction and AGN feedback in TNG and found that, unlike for the less massive halos (M$_{200c} \leq $ 10$^{11.5}$ M$_{\odot}$) which tend to be more gas-rich, the CGM is substantially blown away when TNG halos enter the MW-mass regime (M$_{200c} \simeq$ 10$^{12.0-12.5}$ M$_{\odot}$) wherein, their AGN feedback is mostly kinetic. 
Note that the halos in log M$_{\rm *C}$ = 10.160-10.405 and log M$_{\rm *C}$ = 10.453-10.715 are less massive while those in log M$_{\rm *C}$ = 10.931-11.288 bin lie in the MW-mass category\footnote{As highlighted in §\ref{section: SampSelection}, all of our centrals have M$_{*} > $ 7 $\times$ 10$^9$ M$_{\odot}$. Following Fig. 5 from \citet{weinberger10.1093/mnras/stw2944}, it is well established that every TNG galaxy having a stellar mass M$_{*} > $ 10$^9$ M$_{\odot}$ hosts an SMBH (M$_{\rm bh} > $ 10$^6$ M$_{\odot}$). The presence of an SMBH for all our halos is confirmed in Figures \ref{fig:LMBScatter} - \ref{fig:HMBScatter}.}. Thus, it is important to at least qualitatively discuss the implications of above findings for the present study. CGM expulsion in MW-mass halos naturally implies a drop in the CGM densities. 

A reduced CGM density may make satellite stripping in MW-mass halos more difficult. On the other hand, winds from AGN feedback may enhance satellite stripping by increasing the differential velocity between the satellite and the surrounding medium \citep{nelson2019MNRAS.490.3234N}. Once gas is removed from a satellite it can survive as a cold cloud or heat into the ambient CGM.  A low-density CGM evacuated by AGN feedback will have longer cooling timescales, inhibiting cloud growth. However, if the AGN winds are more metal-enriched (i.e., having an enhanced wind metal loading factor), they can end up increasing the efficiency of metal-line cooling in CGM, thus helping the warmer gas condense into cold phase faster \citep{suresh2015MNRAS.448..895S}.  

We conclude that in general, AGN feedback in TNG should not be a significant source of cold CGM gas. Unlike stellar feedback, because AGN feedback has a long-range effect on the CGM, it could impact the survival of cold gas brought in by satellites. However, given the many possible competing effects, this topic requires further study.   

\subsection{Implications in context of the Local Group}
\label{section: implications}

Considering that many of our galaxies lie in the LMC and MW-mass range, it is worthwhile to view our findings from the perspective of the LG CGM. It is now observationally as well as theoretically well-established that isolated dwarf galaxies generally contain more gas and are more actively star-forming than their counterparts lying within the virial influence of MW-M31 pair \citep{McConnachie_2012, fitts10.1093/mnras/stx1757}. \citet{Putman_2021} inferred that the presence of a diffuse halo environment (such as our LG) is crucial in ensuring the depletion of a satellite's gas reservoir and its subsequent quenching. Quenching becomes progressively less efficient as we move from the least massive (M$_*$ $<$ 10$^6$--10$^7$ M$_{\odot}$) to more massive (M$_*$ $\simeq$ 10$^7$--10$^8$ M$_{\odot}$) to the most massive satellites (M$_*$ $\simeq$ 10$^8$--10$^{10}$ M$_{\odot}$). Multiple simulation-based studies echo similar results \citep{wetzel2015ApJ...807...49W,simpson2018MNRAS.478..548S,kimmel10.1093/mnras/stz2507,Akins_2021,Karunakaran_2021,font10.1093/mnras/stab1332,joshi10.1093/mnras/stab2573}. 

Assuming that the presence of a cold gas reservoir associated with a satellite is a robust indicator of whether or not it is quenched, our findings align with the above statements, since we do find negligible cold gas associated with the least massive satellites while an appreciable amount of cold gas is generally found to be associated with the more massive ones. Furthermore, there exists observational evidence that cold gas streams can be formed during the passage of gas-rich massive satellite galaxies (e.g., the Magellanic Clouds in the Milky Way halo) through the host galaxy's CGM and that these streams (in case of the Milky Way: the Magellanic Stream and potentially other high-velocity clouds; \citet{Richter_2017}) could act as a fodder for future star formation \citep{bekki10.1111/j.1745-3933.2007.00357.x,besla10.1111/j.1365-2966.2012.20466.x,griffiths2018NatAs...2..901M} as well as add substantially to the absorption cross-section of CGM in background quasar spectra.

\section{Summary}
\label{section: Summary}

We have probed the connection between satellite galaxies and their corresponding centrals in the TNG50 simulations from the perspective of the effects on the extended cold CGMs of the host galaxies. Our final sample of 197 halos ranges from sub-MW mass to MW-mass galaxies and spans two orders of magnitude in halo mass. Massive haloes tend to have more cold gas and more satellite galaxies and thus subsequent analysis on a sample with such a broad mass range might give us a highly biased picture about the effects of satellite population on the cold CGM. Therefore, we split our dataset into three mass bins, log M$_{\rm *C}$ = 10.160-10.405, log M$_{\rm *C}$ = 10.453-10.715 and log M$_{\rm *C}$ = 10.931-11.288, such that the total number of satellites, N$_{\rm sats}$, does not correlate with the stellar mass of its associated central, M$_{\rm *C}$. 
The effect on the cold gas reservoir of a halo is studied from the point of view of three different global satellite properties-- N$_{\rm sats}$ (total number of satellites), M$_{\rm tms}$ (total mass in satellites) and M$_{\rm *mms}$ (stellar mass of the most massive satellite). We summarize our findings below-- 

\begin{itemize}
    \item Satellites, on the whole, contribute to the extended radial cold gas densities of their respective hosts, across all mass bins (Fig. \ref{fig:DensProfiles}). However, much of the cold gas mass in galactic halos lies away from the satellites (compare the top and bottom panels in Figs. \ref{fig:NumSats}, \ref{fig:MassSats} and \ref{fig:MostMassiveSat}).
     \item Among the satellite properties probed, N$_{\rm sats}$ (M$_{\rm *mms}$) emerges as the most impactful parameter for the low (middle) mass halos (Figs. \ref{fig:NumSats} and \ref{fig:MostMassiveSat}). N$_{\rm sats}$ correlates best with M$_{\rm cg}$ for the halos in the high-mass bin as well but considering the low number statistics, we refrain from making any conclusive claims.
     \item The number of massive satellites hosted by a galaxy correlates with the halo cold gas mass in a similar way as the total number of satellites. Massive satellites could, therefore, prove to be an effective observable for identifying halos that are likely to have large cold gas reservoirs (Fig. \ref{fig:massive}).
    \item Most satellites on the low-mass end (M$_{*} \lesssim$ 10$^8$ M$_{\odot}$) are devoid of cold gas and hence, can be inferred to have been completely stripped; while more massive satellites have cold gas associated with them, implying their ability to potentially feed the host CGM with cold clouds for a sustained period of time (Fig. \ref{fig:cgmms}).
    \item While the median trends dictate that the cold gas associated with the satellite population accounts for less than 10\% of the cold gas mass in the host CGM, in some cases, they can account for 60\% or more of the cold gas mass in a halo (Fig. \ref{fig:satgas}).

\end{itemize}

Gas removal from satellite galaxies is one among several other existent processes such as IGM accretion, galactic fountain flows, galactic feedback and mergers. Despite this, and the fact that the total cold gas added to the CGM directly from satellites is likely to be small, we find correlations between the number and mass in satellites and the cold gas mass in the CGM, even finding weak correlations between the number of satellites and the cold CGM mass far from satellites. 
Our findings can be used to plan future observations, as we predict that more satellites will result in more cold gas in the CGM of sub-MW and MW-mass galaxies. Importantly, this holds even when only counting observable satellites, which we define as those with stellar masses greater than 10$^8$ M$_{\odot}$. Observable satellites around a halo may hold a key towards accounting for the cold gas contributions from the satellite population. However, the correlation between the satellite population and the cold CGM gas mass has a large scatter, so we argue that large surveys that connect satellite stellar masses to the cold CGM gas reservoir will be required to observe the relationship. On the other hand, the challenge for simulations is to determine what causes an excess of cold CGM gas in satellite-rich systems far from the satellites by measuring the survival and orbits of gas stripped from satellites and by building up our understanding of the satellite-filamentary accretion connection. 

\section*{Data availability}
The complete IllustrisTNG suite of simulations (including TNG50) are publicly available at \url{https://www.tng-project.org/data} \citep{nelson2019ComAC...6....2N}. The scripts and plots for this article can be shared on reasonable request to the corresponding author.

\section*{Acknowledgements}
The authors sincerely thank the JupyterLab Workspace service (all the scripts and resultant figures in this article have been developed on this web-based user interface), provided as a part of the public data access within IllustrisTNG. We thereby, also acknowledge the role of High Performance Computing Center Stuttgart (HLRS) in Germany, the Gauss Centre for Supercomputing (GCS) and Max Planck Computational Data Facility (MPCDF) in running the simulations. MD thanks the CCA at the Flatiron Institute for hospitality while a portion of this research was carried out. The Flatiron Institute is a division of the Simons Foundation. The authors thank the anonymous referee for their useful comments which improved the scientific quality of this paper.

\appendix

\section{The Binning Process}
\label{section: binning}

To bin our galaxies into smaller mass groups, we iteratively select stellar mass ranges and use the 2-sample Kolmogorov-Smirnov (K-S) test \citep{1573387449623294976,smirnov10.1214/aoms/1177730256} to determine how the stellar mass distribution differs when comparing halos with high (satellite-rich) versus low (satellite-poor) numbers of satellites. The 2-sample K-S test has been used as an effective statistical tool in determining the likelihood of two distributions stemming from the same sample \citep{peacock10.1093/mnras/202.3.615,fasano10.1093/mnras/225.1.155,babu2006ASPC..351..127B}. 

To guide the reader, the K-S test comparing the M$_{\rm *C}$ distributions in the top and bottom quartiles (where quartiles refer to the total number of satellites) for the whole sample of 234 halos gives a p-value of 5.516 $\times$ 10$^{-23}$, well below any reasonable threshold that would indicate the stellar masses are drawn from the same distribution. This necessitates the need for a series of iterations which ultimately produces a low mass bin (log M$_{\rm *C}$ = 10.160-10.405), a middle mass bin (log M$_{\rm *C}$ = 10.453-10.715) and a high mass bin (log M$_{\rm *C}$ = 10.931-11.288) such that the stellar mass distributions of their Centrals are roughly similar between the bin-wise bottom and top quartiles (i.e. bottom and top percentile of the total number of satellites). The low (high) mass bin is characterized by the lower 49\% (upper 9\%) of our total halo population. We present the D-statistic and p-values for our final set of bins in Table~\ref{table:KStest}. Note that the total number of halos in the three mass bins is less than the size of our original sample. This is because the p-value for the two bins containing these remaining halos (log M$_{\rm *C}$ = 10.405-10.453 and log M$_{\rm *C}$ = 10.715-10.931) is low, indicating that there is still an undesirable correlation between N$_{\rm sats}$ and M$_{\rm *C}$. Therefore, our analysis omits a total of 37 halos falling in these two mass bins. 


 
Unlike the log M$_{\rm *C}$ = 10.160-10.405 and log M$_{\rm *C}$ = 10.453-10.715 bins, the combined number of halos in the extreme quartile ends of the log M$_{\rm *C}$ = 10.931-11.288 bin (10 halos)  is quite low (see Table \ref{table:FinalSamp}). The reliability of the K-S test is known to reduce significantly in the event of small sample sizes. In such cases, the median and mean absolute deviation (MAD) values are more reliable. Thus, as a final check, we also obtain the median values of M$_{\rm *C}$ ($\bar X$) and the MADs for the extreme mass bin quartiles. The MADs are computed as the mean of $\vert X_{i} - \bar X\vert$, where, $\bar X$ is the median of $X_i$. If the $\bar X \pm$ MADs between quartiles of respective bins overlap, it indicates that the two distributions are similar. These values (see Table~\ref{table:KStest}) show a good overlap in the low and middle mass bins, and indicate that we should treat the high mass bin with skepticism, as we do throughout the paper. We also compute corresponding inter-bin parameter values for the two sets of quartiles for other halo attributes such as M$_{\rm 200c}$, M$_{\rm gas}$, M$_{\rm cg}$, M$_{\rm bh}$ and SFR for the sake of completeness (summarized in Table \ref{table:MAD}).

\begin{figure*}
\centering
\includegraphics[width = \linewidth]{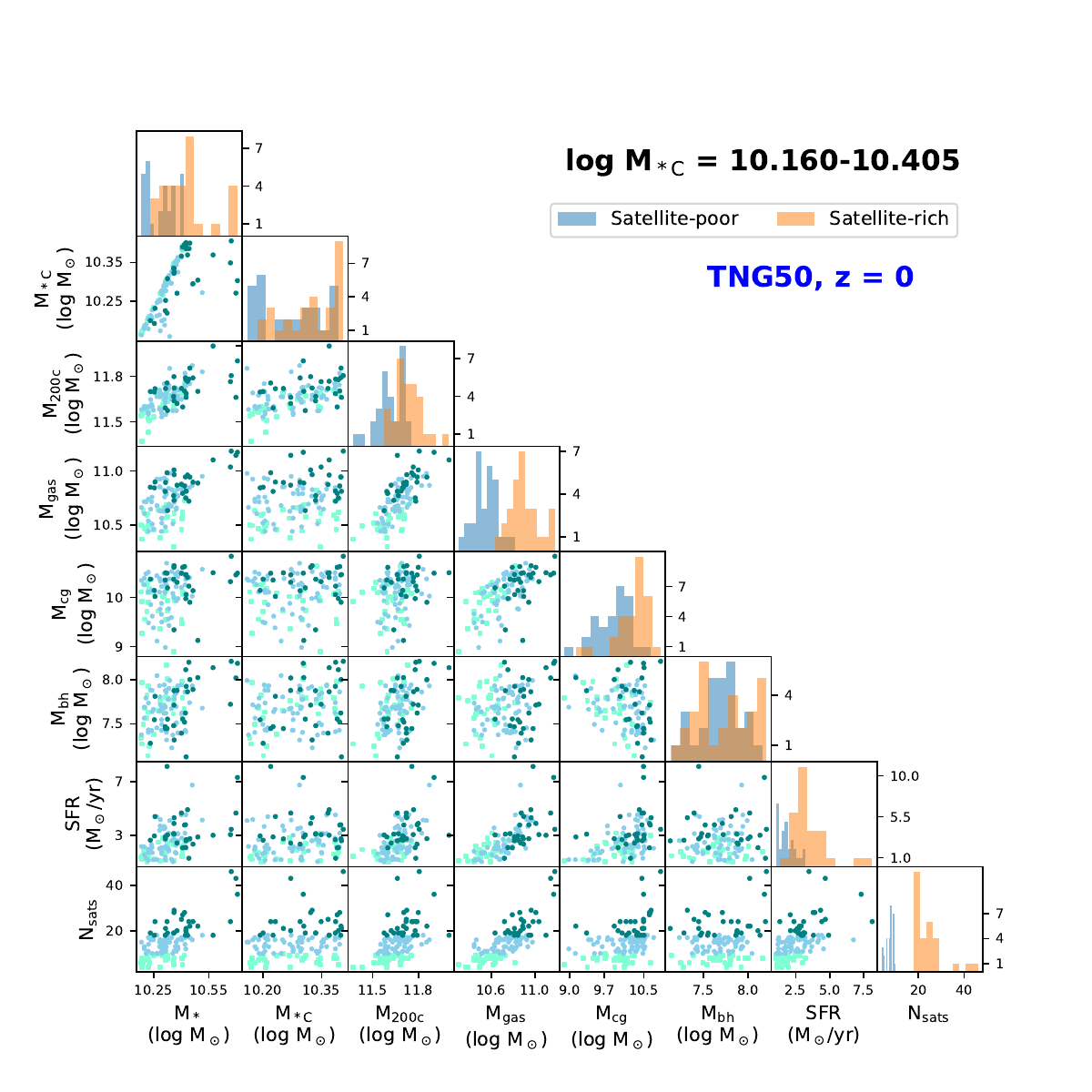}
\caption{Scatter plots for the low mass bin (log M$_{\rm *C}$ = 10.160-10.405) halos probing correlations between Number of satellite galaxies per halo (N$_{\rm sats}$), M$_{*}$, M$_{\rm *C}$, M$_{200c}$, M$_{\rm gas}$, M$_{\rm cg}$, M$_{\rm bh}$ (all in log M$_{\odot}$), and SFR (in M$_{\odot}$ yr$^{-1}$). 25th (Satellite-poor), 50th and 75th (Satellite-rich) quartiles are shown as aqua, blue and teal points respectively. Histograms for the satellite-poor and satellite-rich quartiles for corresponding halo attributes are also shown in the form of blue and orange bars respectively.}
\label{fig:LMBScatter}
\end{figure*}

\begin{figure*}
\centering
\includegraphics[width = \linewidth]{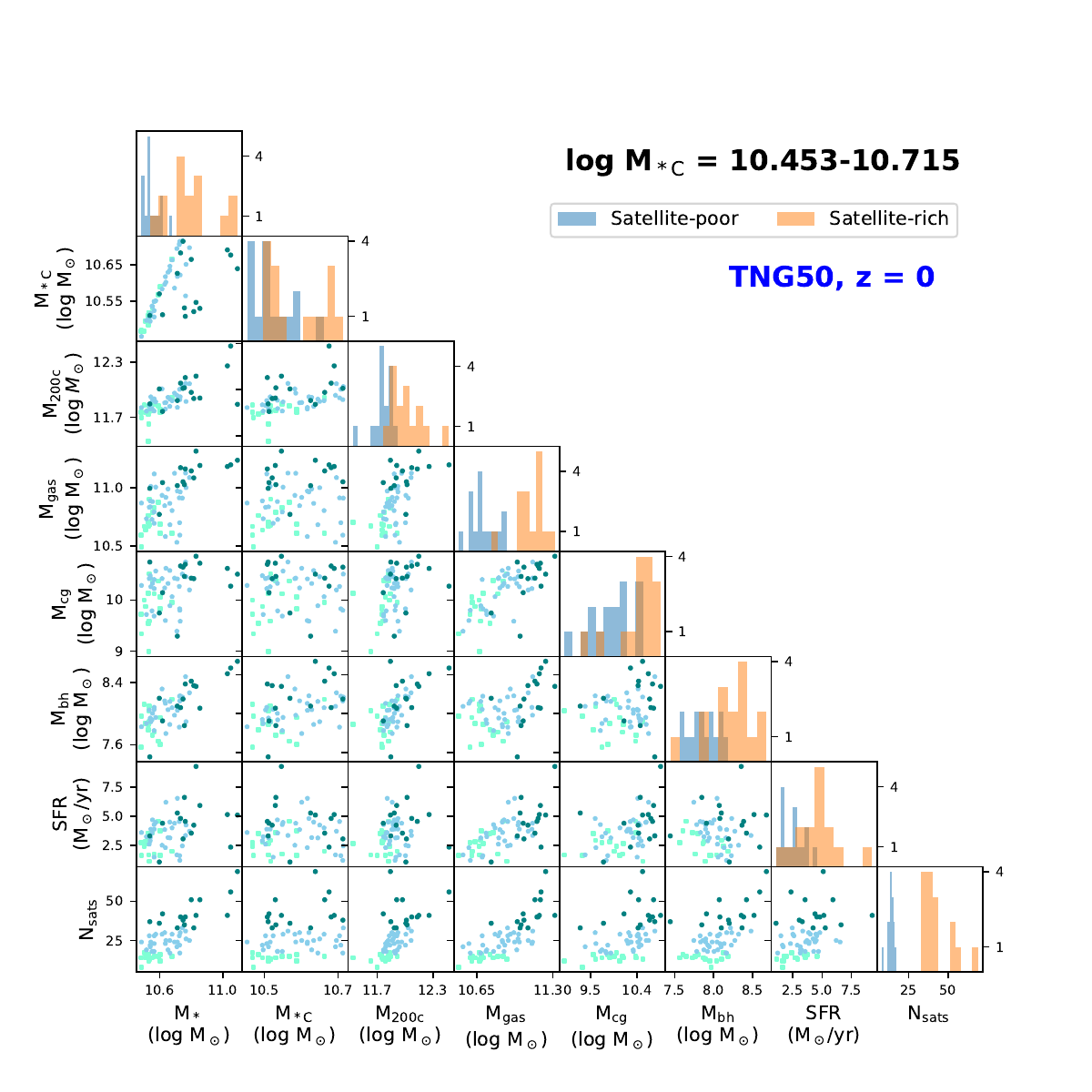}
\caption{Same as Fig.~\ref{fig:LMBScatter}, but for the middle mass (log M$_{\rm *C}$ = 10.453-10.715) bin halos.}
\label{fig:MMBScatter}
\end{figure*}

\begin{figure*}
\centering
\includegraphics[width = \linewidth]{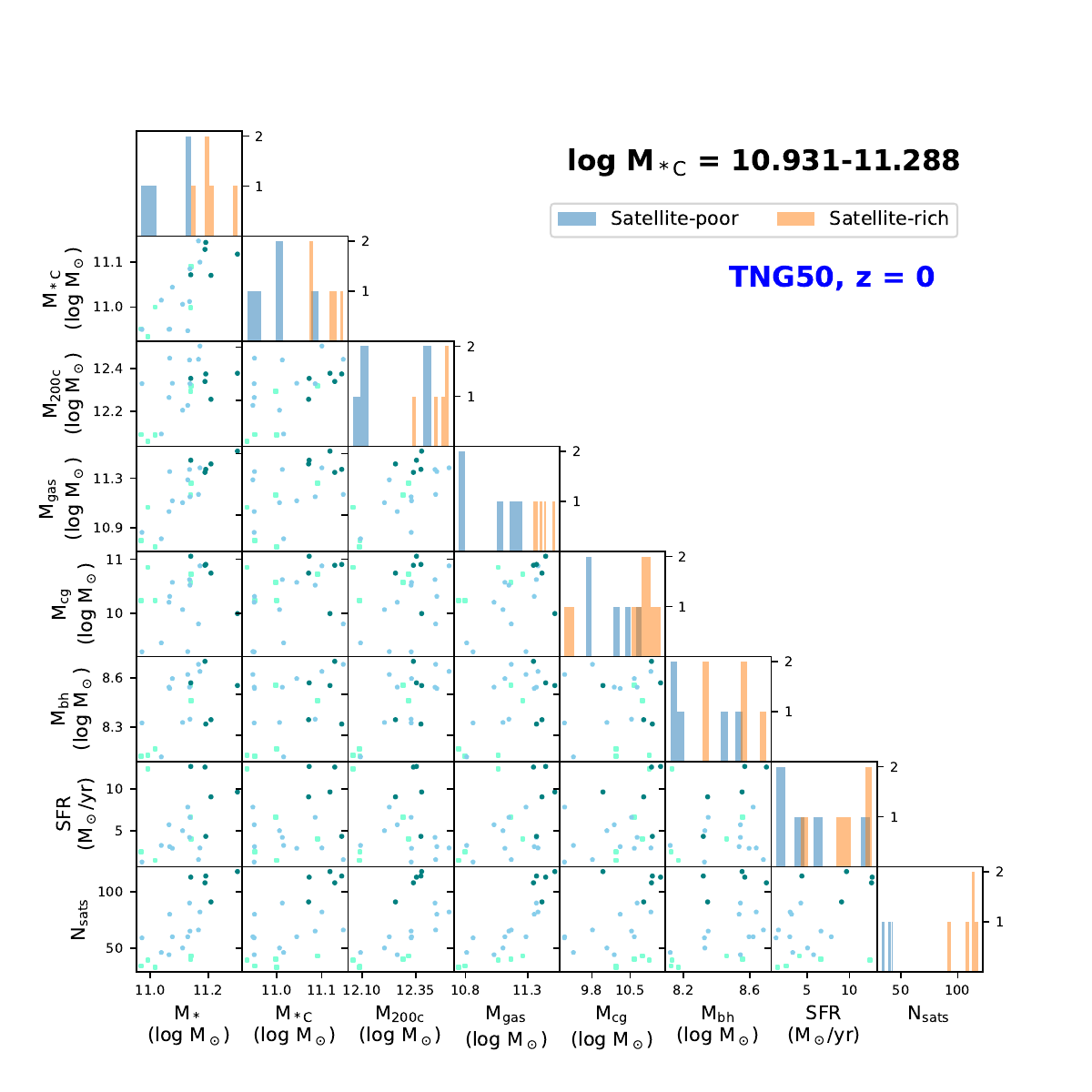}
\caption{Same as Fig.~\ref{fig:LMBScatter}, but for the high mass (log M$_{\rm *C}$ = 10.931-11.288) bin halos.}
\label{fig:HMBScatter}
\end{figure*}

\subsection{Scatter Plots}
\label{section:final}
Our K-S test-based binning gives us mass bins that satisfy the condition of independence of M$_{\rm *C}$ from N$_{\rm sats}$. However, possible inter-dependencies among other halo attributes like gas mass, halo mass, stellar mass, SFR, etc. may later be falsely interpreted purely as a signature of the satellite properties. Therefore, we generate detailed scatter plots for each of our mass bins (log M$_{\rm *C}$ = 10.160-10.405, log M$_{\rm *C}$ = 10.453-10.715 and log M$_{\rm *C}$ = 10.931-11.288 in Figs. \ref{fig:LMBScatter}, \ref{fig:MMBScatter} and \ref{fig:HMBScatter} respectively) involving various group halo attributes like the stellar masses, halo masses, gas masses, black hole masses, cold gas masses, star formation rates and number of satellites. The 25th (low number of satellites; satellite-poor), 50th and 75th (high number of satellites; satellite-rich) quartiles are shown in aqua, blue and teal, respectively. We also plot histograms for the satellite-poor (blue bars) and satellite-rich (orange bars) quartiles.

The correlations between halo attributes are generally similar for the different mass bins. Note that all the mass attributes in Figs.~\ref{fig:LMBScatter}-\ref{fig:HMBScatter} (except the stellar mass of the Central) correspond to the entire group/halo; M$_{\rm gas}$ and M$_{\rm cg}$ refer to the total gas and cold gas in the ISM and CGM of the halo in both central and satellite population while the SFR refers to the total group/halo SFR\footnote{This is the reason why some of the SFRs in Figs.~\ref{fig:LMBScatter}-\ref{fig:HMBScatter} are expectedly higher than the SFR values imposed during sample selection in Sec. \ref{section: SampSelection} - they include the SFR of the satellite galaxies as well as the central.}. Large scatter is seen across all halo attributes irrespective of the mass bin\footnote{We see less pronounced scatter for M$_{\rm 200c}$-M$_{*}$ and M$_{\rm 200c}$-M$_{\rm gas}$, since we expect a more massive halo to generally have a correspondingly larger reserve of stellar and gas mass as compared to a less massive halo.}. Co-variances for halos at the low-end and high-end of each bin also blend-in rather smoothly and do not show any sudden jumps\footnote{The co-variance between N$_{\rm sats}$ and remaining attributes expectedly shows a bi-modal character since we have a clear demarcation between N$_{\rm sats}$ values for the low and high-end for each bin.}. In log M$_{\rm *C}$ = 10.160-10.405 bin (Fig.~\ref{fig:LMBScatter}), as one might expect, we see correlations between M$_{*}$ and M$_{\rm 200c}$, M$_{\rm 200c}$ and M$_{\rm gas}$, and M$_{\rm gas}$ and M$_{\rm cg}$. SFR shows positive trends with M$_{\rm 200c}$, M$_{\rm gas}$ and M$_{\rm cg}$. Finally, it appears by eye that N$_{\rm sats}$ may be correlated with M$_{200c}$ and M$_{\rm gas}$; the difference between the histograms as well as the overlap between the $\bar X \pm$ MAD values (as shown in the Table \ref{table:MAD}) for the corresponding satellite-poor and satellite-rich quartiles further supports this inference. The log M$_{\rm *C}$ = 10.453-10.715 bin halos also show all the above stated correlations except that N$_{\rm sats}$ seemingly shows stronger correspondence with more halo attributes (i.e. M$_{*}$, M$_{\rm 200c}$, M$_{\rm gas}$ and M$_{\rm bh}$) as compared to that for log M$_{\rm *C}$ = 10.160-10.405 halos. The high mass (log M$_{\rm *C}$ = 10.931-11.288) sample contains few halos compared to the other two mass bins, and therefore, the scatter is more pronounced. 
These plots demonstrate that there is no correlation seen between M$_{\rm cg}$ and M$_{\rm *C}$ across all three bins. 

\begin{table}[h]
\caption{The 2-sample K-S test results (D-statistic and p-value), performed between the bottom (BQ) and top quartiles (TQ) of each respective bin, corresponding to our most optimal bin limits. Corresponding percentile as well as the M$_{\rm *C}$ ranges are also stated. Last two columns provide the $\bar X$ $\pm$ MAD values for the respective quartiles per bin. Excellent agreement between these two sets of values confirms the earlier conclusion from the K-S test about the high likelihood of the underlying quartile samples having a common distribution.}
\label{table:KStest}
\begin{footnotesize}
\hspace{-0.4cm}
\begin{minipage}{\linewidth} \centering
\begin{tabular}{l|c|c|cc|c|c|c|c|c|c}
\hline
\hline
{Samples} & {Percentiles} & {Bin limits} & \multicolumn{2}{c|}{2-sample K-S test} & log Med$_{BQ}$ & MAD$_{BQ}$ & log Med$_{TQ}$ & MAD$_{TQ}$ & log ($\bar X_{BQ}$ + & log ($\bar X_{TQ}$ -\\(BQ-TQ) &
& log M$_{\rm *C}\ ({\rm M}_{\odot})$ & D-statistic & p-value & ($\bar X_{BQ}$) & (log) & ($\bar X_{TQ}$) & (log) &  MAD$_{BQ}$) &  MAD$_{TQ}$)\\
\hline
\hline
Low mass & 0--49 & 10.160--10.405 & 0.31 & 0.123 & 10.270 & 9.470 & 10.335 & 9.428 & 10.334 & 10.277\\
Middle mass & 57--83 & 10.453--10.715 & 0.40 & 0.184 & 10.517 & 9.503 & 10.569 & 9.799 & 10.557 & 10.489\\
High mass & 91-100 & 10.931--11.288 & 0.80 & 0.079 & 10.998 & 9.982 & 11.120 & 9.884 & 11.038 & 11.092\\

\hline
\hline
\end{tabular}
\end{minipage}
\end{footnotesize}
\end{table}

\begin{figure*}
\centering
\includegraphics[width = \linewidth]{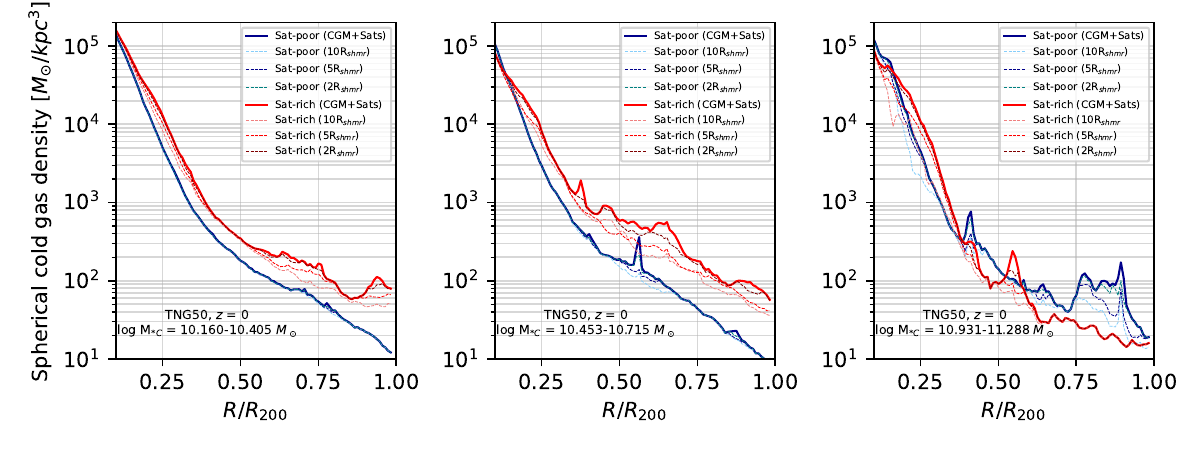}
\caption{Same as Fig. \ref{fig:DensProfiles} but now also with two other radial cuts for the satellite populations. This time along with plotting all cold gas
at r $\gtrsim$ 10.0 $\cdot$ R$_{0.5}$ (i.e. \textit{CGM-Satellites} case; \textit{dashed light blue and light red}) of each satellite within a halo, we also show the mean cold gas spherical densities for $R_{5}$ (r $\gtrsim$ 5.0 $\cdot$ R$_{0.5}$; \textit{dashed dark blue and dark red}) and $R_{2}$ (r $\gtrsim$ 2.0 $\cdot$ R$_{0.5}$; \textit{dashed teal and maroon}). As before, the density profiles for \textit{CGM+Satellites} case are depicted with solid lines. Thus, the difference between similarly colored dashed lines is an indicator of the cold gas lying solely in the extended environments of satellites.}
\label{fig:DensProfilesComp}
\end{figure*}

\section{Spherical cold gas densities for additional radial cut values}
\label{section: comp_dens}
Fig. \ref{fig:DensProfilesComp} shows the cold gas density profiles for two other radial cuts apart from the one shown in Fig. \ref{fig:DensProfiles}. The basic motivation of this exercise is to justify the choice of our adopted radial cut in Fig. \ref{fig:DensProfiles}. As expected, the $R_{2}$ curves (r $\gtrsim$ 2.0 $\cdot$ R$_{0.5}$) lie closest to the respective \textit{CGM+Satellites} curve, since we exclude only the cold gas lying very close to each satellite rather than its extended environment. Nonetheless, a perfect overlap between these two curves is still missing. Thus, some of the satellite contributions seen in Fig. \ref{fig:DensProfiles} are due to the cold gas residing within the satellite(s). 

There is no clear definition of the outer bounds of a satellite galaxy in TNG. A particle is labelled as a satellite galaxy particle if the SUBFIND algorithm computes it to be gravitationally bound to the satellite. On the other hand, one of the parameters observers often resort to, in order to define an adhoc boundary of a galaxy, is its stellar half mass radius. We, therefore, base our radial cut on this observationally motivated parameter. By adopting a rather generous value (r $\gtrsim$ 10.0 $\cdot$ R$_{0.5}$), we ensure that most of the particles within this spatial cut radius most likely already qualify as being gravitationally bound to the satellite (according to the SUBFIND) and/or are seen as associated with the satellite (as recorded in observations).
\begin{sidewaystable}[h!]
\centering
\caption{The inter-bin $\bar X$ and MAD values for the halos belonging to the bottom and top quartiles for M$_{\rm 200c}$, M$_{\rm gas}$, M$_{\rm bh}$ and SFR. Except for the SFR\\ $\bar X$ $\pm$ MAD values for log M$_{\rm *C}$ = 10.453-10.715 and log M$_{\rm *C}$ = 10.931-11.288 halos, the $\bar X$ $\pm$ MAD values for respective quartiles nicely overlap with each other for every other halo attribute. Such an overlap is indicative of a similar distribution for the corresponding halo attribute within a bin.}
\label{table:MAD}
\vspace{2em}
\begin{minipage}{1.09\textwidth} \centering
\hspace{-7cm}
\renewcommand\arraystretch{2} 
  \centering
    \resizebox{\textwidth}{!}
{\begin{tabular}{|l|cc|cc|cc|cc|cc|cc|cc}
\hline
\hline
{\Large Mass bins} & \multicolumn{2}{c|}{\Large log M$_{\rm 200c}$ (M$_{\odot}$)} & \multicolumn{2}{c|}{\Large log M$_{\rm gas}$ (M$_{\odot}$)} & \multicolumn{2}{c|}{\Large log M$_{\rm bh}$ (M$_{\odot}$)} & \multicolumn{2}{c|}{\Large SFR (M$_{\odot}$ yr$^{-1}$)} & \multicolumn{2}{c|}{\Large log M$_{\rm cg}$ (M$_{\odot}$)}\\ &
$\bar X_{BQ}$ (MAD$_{BQ}$) & $\bar X_{TQ}$ (MAD$_{TQ}$) & $\bar X_{BQ}$ (MAD$_{BQ}$) & $\bar X_{TQ}$ (MAD$_{TQ}$) & $\bar X_{BQ}$ (MAD$_{BQ}$) & $\bar X_{TQ}$ (MAD$_{TQ}$) & $\bar X_{BQ}$ (MAD$_{BQ}$) & $\bar X_{TQ}$ (MAD$_{TQ}$) & $\bar X_{BQ}$ (MAD$_{BQ}$) & $\bar X_{TQ}$ (MAD$_{TQ}$) \\
\hline
\hline
log M$_{\rm *C}$ = 10.160-10.405 & 11.612 (10.800) & 11.723 (10.905) & 10.522 (9.818) & 10.824 (10.148) & 7.669 (7.314) & 7.674 (7.415) & 1.654 (0.448) & 2.750 (0.655) & {10.071 (9.897)} & {10.255 (10.030)}\\
log M$_{\rm *C}$ = 10.453-10.715 & 11.772 (10.854) & 12.011 (11.456) & 10.609 (10.010) & 10.936 (10.366) & 7.859 (7.433) & 8.014 (7.662) & 1.892 (0.740) & 2.550 (0.764) & {10.259 (10.152)} & {10.336 (10.021)}\\
log M$_{\rm *C}$ = 10.931-11.288 & 12.093 (11.460) & 12.370 (11.160) & 10.991 (10.453) & 11.219 (10.497) & 8.154 (7.656) & 8.342 (7.968) & 2.267 (1.203) & 3.089 (0.710) & {10.410 (10.217)} & {10.888 (10.390)}\\
\hline
\hline
\end{tabular}}
\end{minipage}
\end{sidewaystable}








\bibliography{aas}{}
\bibliographystyle{aasjournal}


\label{lastpage}
\end{document}